\documentclass[fleqn,usenatbib]{mnras}

\usepackage{amsmath}	
\usepackage{amssymb} 
\usepackage{txfonts}
\usepackage{mathptmx}
\usepackage[T1]{fontenc}
\usepackage{ae,aecompl}

\usepackage{xcolor}
\usepackage{graphicx}	

\newcommand{\bd}[1]{\mbox{\boldmath $#1$}}
\newcommand{\angstrom}{\textup{\AA}}

\title[Optical continuum PRM of Mrk509]{Optical continuum photometric reverberation mapping of the Seyfert-1 galaxy Mrk509}

\author[F. Pozo Nu\~nez et al.]{
F. Pozo Nu\~nez$^{1,2,3}$\thanks{E-mail: francisco.pozon@gmail.com}\href{https://orcid.org/0000-0002-6716-4179}{\includegraphics[scale=0.9]{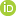}},
N. Gianniotis$^{4}$\href{https://orcid.org/0000-0002-2187-5522}{\includegraphics[scale=0.9]{orcidicon.png}},
J. Blex$^{2}$,
T. Lisow$^{2}$,
R. Chini$^{2,6}$,
\newauthor ~K. L. Polsterer$^{4}$, J.-U. Pott$^{5}$\href{https://orcid.org/0000-0003-4291-2078}{\includegraphics[scale=0.9]{orcidicon.png}}, J. Esser$^{5}$, and G. Pietrzy\'{n}ski$^{3}$ \\
$^{1}$Haifa Research Center for Theoretical Physics and Astrophysics, Haifa 31905, Israel\\
$^{2}$Astronomisches Institut, Ruhr--Universit\"at Bochum, Universit\"atsstra{\ss}e 150, 44801 Bochum, Germany\\
$^{3}$Centrum Astronomiczne im. Mikolaja Kopernika, PAN, Bartycka 18, 00-716 Warsaw, Poland\\
$^{4}$Heidelberg Institute for Theoretical Studies gGmbH, Heidelberg, Germany\\
$^{5}$Max-Planck Institut f\"ur Astronomie, K{\"o}nigstuhl 17 Heidelberg, Germany\\
$^{6}$Instituto de Astronom\'{i}a, Universidad Cat\'{o}lica del Norte, Avenida Angamos 0610, Casilla 1280,Antofagasta, Chile\\
}

\date{Accepted 2019 October 3. Received 2019 October 1; in original form 2019 August 29}

\pubyear{2019}

\begin{document}
\label{firstpage}
\pagerange{\pageref{firstpage}--\pageref{lastpage}}
\maketitle


\begin{abstract}

We present the results of a two year optical continuum photometric reverberation mapping campaign  carried out on the nucleus of the Seyfert-1 galaxy Mrk509. Specially designed narrow-band filters were used in order to mitigate the line and pseudo-continuum contamination of the signal from the broad line region, while allowing for high-accuracy flux-calibration over a large field of view. We obtained light curves with a sub-day time sampling and typical flux uncertainties of $1\%$. The high photometric precision allowed us to measure inter-band continuum time delays of up to $\sim 2$ days across the optical range. The time delays are consistent with the relation $\tau \propto \lambda^{4/3}$ predicted for an optically thick and geometrically thin accretion disk model. The size of the disk is, however, a factor of 1.8 larger than predictions based on the standard thin-disk theory. We argue that, for the particular case of Mrk509, a larger black hole mass due to the unknown geometry scaling factor can reconcile the difference between the observations and 
theory.

\end{abstract}




\section{Introduction}

\begingroup
\setlength{\tabcolsep}{10pt} 
\renewcommand{\arraystretch}{1.5} 
\begin{table*}
\begin{center}
\caption{Characteristics of Mrk509.}
\label{table1}
\begin{tabular}{@{}cccccccc}
\hline\hline

$\alpha$ (2000)$^{(1)}$ & $\delta$ (2000)$^{(1)}$ & $z^{(2)}$ & $D_L^{(2)}$ & $M_{\rm BH}^{(3)}$ & $\tau_{\rm H\beta}^{(3)}$ & $\sigma_{\rm H\beta}^{(3)}$ &
$A_V^{(3)}$ \\
      & & & (Mpc) & ($M_{\odot}$) & (days) & (km s$^{-1}$) & (mag) \\
\hline
20:44:09.7 & -10:43:25.0 & 0.0344 & 145.0 & $14.3 \pm 1.2 \times 10^{7}$ & $79.6^{+6.1}_{-5.4}$ & $1276\pm 28$ & 0.309 \\
\hline
\end{tabular}
\end{center}
\noindent
{\em References:}
[1] - NED database; 
[2] - \cite{1993AJ....105.1637H}; 
[3] - \cite{2004ApJ...613..682P}; The velocity dispersion of the H$\beta$ emission line ($\sigma_{\rm H\beta}$) together with the time delay between the optical continuum and the H$\beta$ emission line ($\tau_{\rm H\beta}$) were used to estimate $M_{\rm BH}$.
\end{table*}
\endgroup

Active galactic nuclei (AGN) are believed to be powered by an accretion disk around a super-massive black hole (SMBH). The strong radiation from the accretion disk photo-ionizes the gas clouds in the broad-line region (BLR) giving rise to the characteristic emission lines observed in the spectrum of quasars and Seyfert galaxies (e.g. \citealt{1979RvMP...51..715D}). The velocity of the BLR clouds combined with its average distance to the accretion disk can be used to estimate the black hole mass in AGN. The BLR responds to the strong and variable UV/Optical continuum at very short  time-scales, and thus at small distances ($\sim$ 1 to 250 light days) from the accretion disk. The small distance means that it is very difficult to resolve the central engine of AGN unless several radio telescopes are combined to create an earth-size detector capable to achieve the high spatial resolution needed (see the reviews by \citealt{2015arXiv150102001A} and \citealt{2015ARA&A..53..365N}). Recent developments in instrumentation allowed to resolve the accretion disk and the BLR system for the particular case of very nearby active galaxies (\citealt{{2018Natur.563..657G}}; \citealt{2019ApJ...875L...4E}). However, it will be impossible to resolve a large sample of more distant AGN in the foreseeable future. Fortunately, in order to estimate black hole masses for a larger sample of objects located at different redshifts, we can resort to the reverberation mapping (RM) method (\citealt{1973ApL....13..165C}; \citealt{1982ApJ...255..419B}; \citealt{1986ApJ...305..175G}). RM is independent of the spatial resolution of the instrument and relies only on the strong, intrinsic variability to measure the time delay, $\tau$, between changes in the accretion disk continuum and the emission lines from the BLR. This allows estimating the average distance of the BLR clouds to the accretion disk ($R_{\rm{BLR}} = c\cdot\tau_{\rm{BLR}}$, $c$ is the speed of light). Through the combination of spectroscopic (e.g. ; \citealt{2004ApJ...613..682P}; \citealt{2012ApJ...755...60G}) and photometric monitoring (\citealt{2011A&A...535A..73H}; \citealt{2012A&A...545A..84P}; \citealt{2012ApJ...747...62C}), the method has revealed the size of the BLR, black hole masses and Eddington ratios in about 100 AGN \citep[e.g.][and references therein]{2014ApJ...782...45D}.

Based on RM measurements of several nearby low-luminosity Seyfert-1 galaxies and a few distant high luminosity quasars, a tight relationship between the accretion disk $5100$\,\AA\ monochromatic luminosity and the size of the BLR has been established ($R_{\rm BLR}\propto L_{\rm{AD}}^{\alpha}$; \citealt{2000ApJ...533..631K}; \citealt{2009ApJ...705..199B}; \citealt{2013ApJ...767..149B}; \citealt{2016ApJ...825..126D}). The radius-luminosity relation has been used to estimate single-epoch black hole masses in larger samples and at different redshifts (e.g. \citealt{2013ApJ...774...67T}; \citealt{2014ApJ...794...77F}; \citealt{2015ApJ...809..123H}; \citealt{2017ApJ...839...93P}). However, it remains unclear what is the physical interplay between the accretion disk and the BLR, and whether BLR continuum contamination could bias the inferred accretion disk optical continuum luminosities (e.g. \citealt{2019NatAs...3..251C}).

Some models suggest that BLR clouds are the consequence of strong dusty wind formed in colder regions of the accretion disk atmosphere (\citealt{2011A&A...525L...8C}). The BLR gas can become exposed to the strong irradiation from the central continuum source as they move further away from the disk surface, hence connecting the outer part of the disk with the inner edge of the hot dust distribution (\citealt{2012MNRAS.426.3086G}; \citealt{2014cosp...40E.597C}; \citealt{2014A&A...561L...8P}; \citealt{2018A&A...620A.137R}). The dusty wind scenario has been supported with a RM campaign of the circumnuclear hot dust in the Seyfert-1 galaxy NGC\,4151 (\citealt{2015A&A...578A..57S}). Recent monitoring of NGC\,4151 by \cite{2019A&A...621A..46E} showed evidence of correlated changes between the dust radius and the shape variations of the Pa$\beta$ BLR emission line, suggesting a common origin for the BLR and the dust clouds which are produced in cooler regions of the accretion disk.

Most of the models assume that AGN have sub-Eddington accretion rates described by the standard thin thermal accretion disk theory (\citealt{1973A&A....24..337S}). While geometrically thin and optically thick disk models have been able to fit the observed spectral energy distribution (SED) in several AGN (e.g. \citealt{2008Natur.454..492K}; \citealt{2015MNRAS.446.3427C}), there are a number of cases where the results are not satisfactory (e.g. \citealt{2007ApJ...668..682D}; \citealt{2014ApJ...783...46K}; \citealt{2016ApJ...818L...1S}).

According to the standard accretion disk theory of \cite{1973A&A....24..337S}, the effective temperature of a thin disk changes with its radius and can be expressed as a function of the black hole mass and accretion rate (e.g. \citealt{2007MNRAS.380..669C}; \citealt{2008ApJ...677..884L}; \citealt{2010ApJ...712.1129M}; \citealt{2016ApJ...821...56F}). The radial extend of the accretion disk can therefore be proved by studying the continuum emission at different wavelengths. Similar to the stratification and reprocessing effects observed in the BLR, the radiation from the innermost part of the accretion disk, closer to the SMBH, has the peak of the emission at shorter wavelengths and its variability is observed with a time delay with respect to the outer and cooler parts of the disk which are traced by longer wavelengths. This effect can be interpreted as the light travel time across the accretion disk (e.g. 
\citealt{1998ApJ...500..162C}). Therefore, time delays between light curves at different continuum bands provide valuable information about the size ($R_{\rm AD}\sim c\cdot\tau_{\rm AD}$) and the temperature stratification across the disk, both crucial parameters to test the standard thin-disk theory in AGN (e.g. \citealt{2008ApJ...677..884L}; \citealt{2013ApJ...772....9C}). 

Accretion disk time delays between the UV-optical bands have been detected for a few AGN over the past years (\citealt{1997ApJS..113...69W}; \citealt{1998ApJ...500..162C}; \citealt{2003ASPC..290..119O}; \citealt{2005ApJ...622..129S}; \citealt{2007MNRAS.380..669C}; \citealt{2015ApJ...806..129E}; \citealt{2016ApJ...821...56F}). The reported uncertainties are large, likely due to under-
sampled light curves. Moreover, a large part of those experiments have been carried out using broad-band filters which can bias the 
results due to the contribution of the BLR emission. In that context, \cite{2017PASP..129i4101P} has recently introduced a photometric 
RM experimental design using a specific set of narrow-band filters with the aim to mitigate the BLR emission-line contamination and 
quantify the effect of the BLR diffuse continuum contribution (\citealt{2019NatAs...3..251C}).

\cite{1998ApJ...500..162C} found that the observed time delays seems to be consistent with the delay-wavelength relation $\tau\propto\lambda^{4/3}$ predicted by geometrically thin accretion disk models.The absolute disk sizes are, however, larger by a factor of $\sim3$ than the expected based on standard thin-disk theory (\citealt{2005ApJ...622..129S}; \citealt{2014ApJ...788...48S}; \citealt{2015ApJ...806..129E}; \citealt{2016ApJ...821...56F}; \citealt{2017ApJ...836..186J}; \citealt{2018ApJ...857...53C}; but see also \citealt{2018ApJ...862..123M}). Interestingly, microlensing studies of luminous lensed quasars have independently reached similar conclusions (\citealt{2006ApJ...648...67P}; \citealt{2007ApJ...661...19P}; \citealt{2010ApJ...712.1129M}; \citealt{2013ApJ...769...53M}; \citealt{2016AN....337..356C}), although the flux ratio of the lensed images are sensitive to the size of the emitting region at a particular wavelength. Moreover, the microlensing technique only allows to study the accretion disk in more distant and high-luminosity quasars, while RM can also study local low-luminosity AGN, especially with smaller telescopes.

\begin{figure*}
  \centering
  \includegraphics[width=\columnwidth]{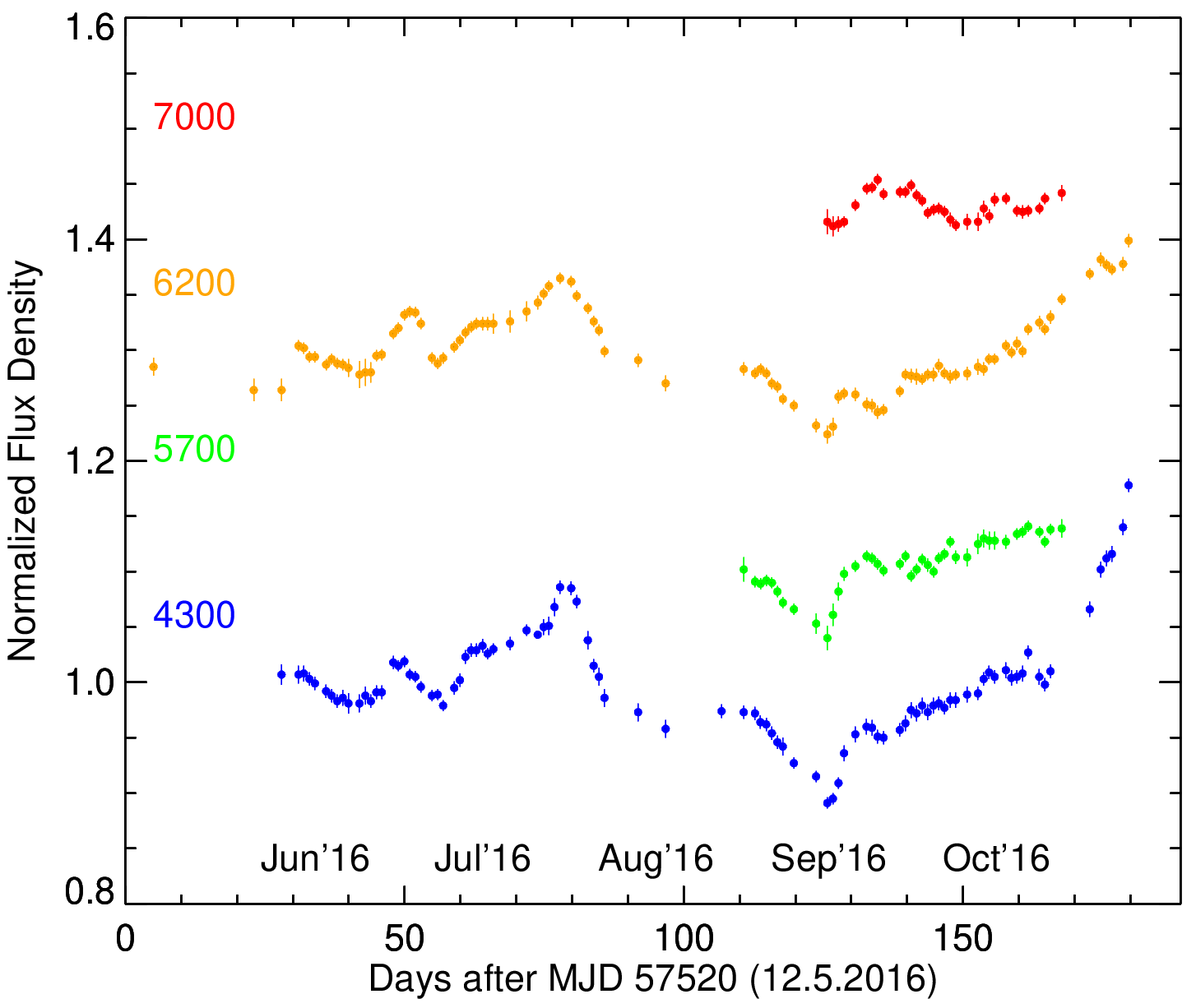}
  \includegraphics[width=\columnwidth]{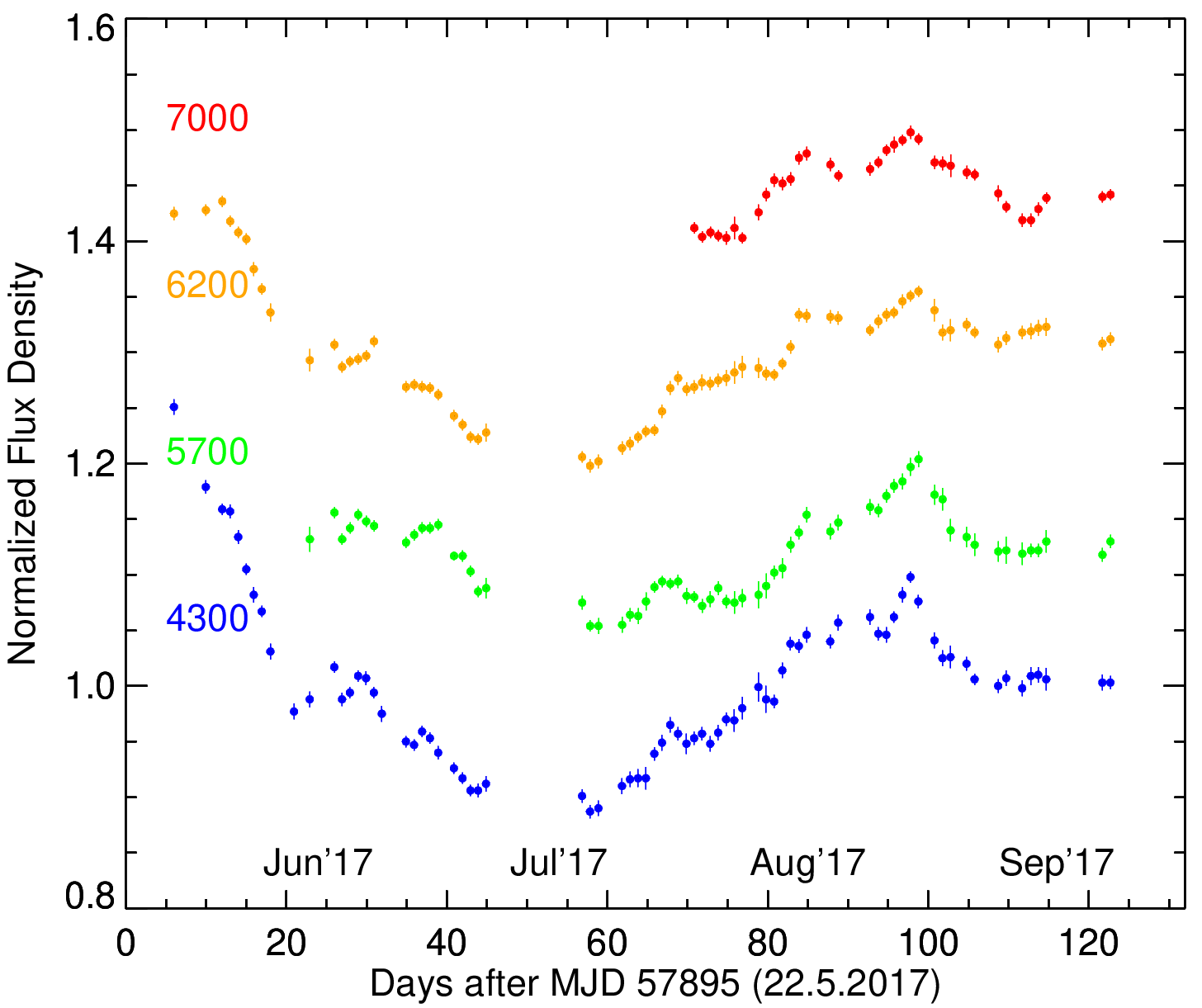}
  \caption{Normalized light curves of Mrk509 for the period between May 2016 and November 2016 ({\it 
  left}) and for the period between May 2017 and September 2017 ({\it right}).
  The light curves are vertically shifted by multiples of 0.2 for clarity.}
\label{lc_opt}
\end{figure*}

Mrk509 is a luminous Seyfert 1 galaxy located at a distance of 145 Mpc and redshift $z = 
0.0344$ (\citealt{1993AJ....105.1637H}). Due to its high brightness, strong variability and characteristic outflows, it has been the target of several X-ray/Optical spectroscopic and photometric monitoring campaigns (e.g., \citealt{2011A&A...534A..36K}; \citealt{2011A&A...534A..39M}; \citealt{2014A&A...567A..44B}). 

Continuum time delays were observed by \cite{2005ApJ...622..129S} who used broad-band $BVRI$ variations and attributed the relative lags between the $B$ and the $VRI$ filters to the light time travel effect and thus to the geometrical size of the region that emits optical continuum.

In this paper, we present the first optical narrow-band continuum photometric reverberation mapping study carried out on the nucleus of Mrk509. We measured continuum time delays using specially designed set of narrow-band filters and discuss the results in the context of emission from an optically thick and geometrically thin accretion disk.

\section{OBSERVATIONS AND DATA REDUCTION}

The photometric monitoring was conducted between May 17 and November 07, 2016, and between May 05 and September 21, 2017, with the robotic 46\,cm telescope of the Wise observatory in Israel. Through Mrk509 redshift of z = 0.034, the narrow-bands $4300\pm50$, $5700\pm50$, $6200\pm60$, and $7000\pm60$\,\AA\, were used to trace the AGN emission line-free continuum variations. An earlier monitoring carried out in 2014 was performed as part of a RM campaign of the BLR using the robotic 15cm VYSOS-6 and 40cm BMT telescopes located at the Bochum Observatory, near Cerro Armazones in Chile (Blex et al. in prep). The Bochum observations were carried out using the broad-band Sloan-$u$ ($3609\pm300$) and narrow-band 680 ($6800\pm120$) which traces the ultra-violet (UV) continuum variations and the H$\alpha$ emission-line response respectively. Below we discuss details and the implications of the BLR observations (Sect. 4). Figure~\ref{fig:filters1} shows the position of the narrow-band filters together with the spectrum of Mrk509 obtained from the AGN Watch monitoring database (\citealt{1996ApJ...471..737C}). The characteristics of Mrk509 are summarized in Table~\ref{table1}. 

The images were reduced following standard procedures for image reduction, including bias, dark current, flatfield, astrometry and astrometric distortion corrections performed with IRAF\footnote{IRAF is distributed by the National Optical Astronomy Observatory, which is operated by the Association of Universities for Research in Astronomy (AURA) under cooperative agreement with the National Science Foundation.} packages and custom written tools, in combination with SWARP (\citealt{2002ASPC..281..228B}), SCAMP (\citealt{2006ASPC..351..112B}), and Astrometry.net (\citealt{2010AJ....139.1782L}) routines. A more detailed description of the filters, and data reduction can be found in \cite{2017PASP..129i4101P} for the Wise 46\,cm telescope, and in \cite{2015A&A...576A..73P} for the Bochum VYSOS-6 and BMT telescopes.

\begingroup
\setlength{\tabcolsep}{4.5pt} 
\renewcommand{\arraystretch}{1.5} 
\begin{table}
\begin{center}
\caption{Photometric observations.}
\label{sumphotres}
\begin{tabular}{@{}cccc}
\hline\hline
Filter & $\lambda_{eff}^{1}$ & $\rm F_{Total}^{2}$ & No. of observations \\
       &   (\AA)             & 2016/2017 (mJy)       &    2016/2017           \\   
\hline
 NB$4300$       &  $4311$      & $17.11\pm0.11$/$17.59\pm0.12$ & $96$/$76$          \\
 NB$5700$       &  $5688$      & $18.21\pm0.10$/$17.41\pm0.11$ & $42$/$65$          \\
 NB$6200$       &  $6208$      & $21.03\pm0.12$/$21.37\pm0.13$ & $97$/$74$          \\
 NB$7000$       &  $7018$      & $22.69\pm0.13$/$23.14\pm0.13$ & $32$/$36$          \\
\hline
\end{tabular}
\end{center}
\noindent
$^{1}$Effective central wavelength: $\int \lambda T(\lambda)d\lambda/ 
T(\lambda)d\lambda$ where $\lambda$ is the wavelength and $T$ the filter transmission.\\
$^{2}$ $\rm F_{Total}$ refer to the mean of the total flux ranges during our 
monitoring. Fluxes are corrected by galactic foreground extinction.
\end{table}
\endgroup

\subsection{Light Curves}
\label{sec:lightcurves} 

The light curves were extracted using image subtraction techniques based on the algorithms implemented in the ISIS package (\citealt{1998ApJ...503..325A}; \citealt{2000A&AS..144..363A}). 
The image subtraction procedure together with a comparison with traditional aperture photometry is explained in detail by \cite{2017PASP..129i4101P}; here we describe only its main characteristics. First, we construct a reference frame by co-adding the images with the best quality. Then, the reference frame is convolved with a spatially variable kernel to match the point-spread function (PSF) of each individual frame. The convolved reference frame is subtracted from the individual images in order to isolate the AGN variable flux. The final step is the extraction of the nuclear flux which is performed on the resulting difference images using PSF photometry. The quality achieved in the subtracted images allows us to measure the nuclear flux of the AGN with a photometric precision of $0.5\%-1.0\%$. We also used traditional aperture photometry on the original images in order to compare the performance of both methods. Special care was taken in selecting the aperture that maximizes the signal-to-noise ratio (S/N) and minimize the contribution of the host galaxy. The photometric precision obtained from the aperture photometry is $1.2\%-2.0\%$. The performance of image subtraction strongly correlates with the quality of the PSF model kernel (\citealt{2017PASP..129i4101P}), and which in turns depends on the amount of the stars in the field. The field of Mrk509 contains $\sim5000$ stars, which is considered a crowded field comparing with Seyfert-1 galaxies located at similar redshifts. This makes image subtraction to outperform aperture photometry in this particular case.
The differential fluxes obtained from the image subtraction process are converted to flux units by performing aperture photometry on the reference frame. We find that an aperture of 6.0 arcsec maximizes the S/N and delivered the lowest absolute scatter for the fluxes.

The absolute flux calibration was obtained using the measured fluxes of reference stars from \cite{2009AJ....137.4186L} observed on the same nights as Mrk509, considering the atmospheric extinction at the Wise observatory and the re-calibrated galactic foreground extinction values presented by \cite{2011ApJ...737..103S}. Based on high-resolution stellar templates of our standard stars, we selected only the stars that have moderate absorption around the filter bandpasses. We expect that any residual over-estimation of the flux in the bands is $\leq10\%$. A summary of the photometric results in all bands are listed in Table~\ref{sumphotres}. The normalized light curves for campaigns 2016 and 2017 are shown in Figure~\ref{lc_opt}. The fluxes in all bands are given in tables A1 and A2 in the Appendix.

\begingroup
\setlength{\tabcolsep}{4.5pt} 
\renewcommand{\arraystretch}{1.5} 
\begin{table}
\begin{center}
\caption{Host galaxy and AGN optical fluxes for 2016 and 2017 campaigns.}
\label{table3}
\begin{tabular}{@{}ccc}
\hline\hline
Filter & Galaxy & AGN$^{1}$ \\
       & 2016/2017 (mJy) & 2016/2017 (mJy) \\   
\hline
NB$4300$ & $3.12\pm0.72$/$2.94\pm0.55$ & $13.93\pm0.86$/$14.73\pm0.75$ \\
NB$5700$ & $7.63\pm0.74$/$7.41\pm0.56$ & $10.62\pm0.74$/$10.02\pm0.76$ \\
NB$6200$ & $7.93\pm0.77$/$7.64\pm0.81$ & $13.10\pm0.88$/$13.73\pm0.91$ \\
NB$7000$ & $10.52\pm3.01$/$7.81\pm0.86$ & $12.20\pm1.73$/$15.31\pm0.94$ \\
\hline
\end{tabular}
\end{center}
\end{table}
\endgroup

\subsection{Host subtracted AGN luminosity and nuclear reddening}
\label{sec:agnlumino}

\begingroup
\setlength{\tabcolsep}{4.5pt} 
\renewcommand{\arraystretch}{1.5} 
\begin{table*}
\begin{center}
\caption{Observed-frame inter-band continuum time delays for 2016 and 2017 campaigns.}
\label{tabledelay}
\begin{tabular}{@{}ccccc}
\hline\hline
Filter & ICCF       & DCF        & $\mathcal{V_{N}}$ \\
       & 2016/2017  & 2016/2017  & 2016/2017 \\   
       & (days)     & (days)     & (days) \\
\hline
NB$4300$ & $0.00^{+0.20}_{-0.20}[1.00]$/$0.00^{+0.20}_{-0.31}[1.00]$ & $0.00^{+0.21}_{-0.20}[1.00]$/$0.00^{+0.23}_{-0.30}[1.00]$ & $0.01^{+0.21}_{-0.19}$/$0.00^{+0.21}_{-0.33}$ \\
NB$5700$ & $0.92^{+0.11}_{-0.36}[0.77]$/$1.12^{+0.82}_{-0.92}[0.82]$ & $0.95^{+0.10}_{-0.40}[0.77]$/$1.10^{+0.81}_{-1.03}[0.82]$ & $0.96^{+0.18}_{-0.42}$/$1.03^{+0.98}_{-1.01}$ \\
NB$6200$ & $1.90^{+0.71}_{-0.72}[0.89]$/$1.72^{+0.12}_{-0.56}[0.93]$ & $1.81^{+0.70}_{-0.73}[0.88]$/$1.72^{+0.12}_{-0.61}[0.93]$ & $1.82^{+1.02}_{-0.57}$/$1.71^{+0.49}_{-0.52}$ \\
NB$7000$ & $1.89^{+1.20}_{-1.22}[0.73]$/$2.11^{+0.61}_{-0.46}[0.74]$ & $1.99^{+1.19}_{-1.21}[0.75]$/$2.10^{+0.60}_{-0.41}[0.74]$ & $2.10^{+1.12}_{-0.96}$/$2.01^{+0.48}_{-0.28}$ \\
\hline
\end{tabular}
\end{center}
\noindent
{\em Notes:}
The maximum correlation coefficient $R_{max}$ is given in parenthesis for both ICCF and DCF methods.
\end{table*}
\endgroup

To disentangle the host and AGN contributions to the total flux in the bands, we used the flux variation gradient (FVG) method (\citealt{1981AcA....31..293C}; \citealt{2004MNRAS.350.1049G}; \citealt{2010ApJ...711..461S}; \citealt{2014A&A...561L...8P}). In brief, the total fluxes obtained through different bands and same apertures follow a linear slope representing the AGN color, while the slope of the host galaxy contribution lies in a well defined range (\citealt{2010ApJ...711..461S}). The AGN slope is determined
through a bisector linear regression analysis (\citealt{1990ApJ...364..104I}). Averaging over the intersection area between the AGN and the host galaxy slopes yields the host galaxy contribution at the time of the monitoring campaign. The FVG diagrams are shown in Figure~\ref{fvg} in the Appendix. The bisector linear regression yields a linear gradient of $\Gamma\sim1$ during both 2016 and 2017 campaigns. The results are consistent with the gradients obtained for other Seyfert-1 galaxies (\citealt{1992MNRAS.257..659W}; \citealt{2010ApJ...711..461S}). Through the use of high-resolution Hubble Space Telescope images, \cite{2009ApJ...705..199B} performed the modeling of the host galaxy profile in Mrk\,509 and found a bulge morphology type. The host galaxy spectral energy distribution (SED) obtained from the FVG analysis is consistent, within the error margins, with a host bulge model spectrum (Appendix Figure~\ref{sed}).

An important point to consider when isolating the true SED of an AGN is the internal AGN reddening (\citealt{2004ApJ...616..147G}; \citealt{2007arXiv0711.1013G}). Neglecting the effect of nuclear extinction can result in luminosities being underestimated up to a factor of 4 and 10 in the optical and UV respectively (\citealt{2017MNRAS.467..226G}). Here, we estimate the nuclear reddening and extinction of Mrk509 directly from the FVG analysis. The bisector method yields a linear gradient of $\Gamma_{4300-5700} = 1.14\pm0.06$, by assuming the unreddened or intrinsic color of AGN to be $B-V = 0.0$ ($\Gamma_{BV} = 1.10$ or $\Gamma_{4300-5700} = 1.18$, \citealt{1992MNRAS.257..659W}; \citealt{1997MNRAS.292..273W}) we find a nuclear reddening E$(4300-5700) = 0.032$, consistent with E$(B-V) = 0.0\pm0.02$ found by \cite{1992MNRAS.257..659W} and \cite{1997MNRAS.292..273W} for Mrk509. Considering the AGN reddening curve of \cite{2004ApJ...616..147G}, the nuclear reddening E$(4300-5700) = 0.032$ corresponds to a visual extinction $A_{v}\sim0.15$ mag. The intrinsic SED for the nuclear region, after accounting for host galaxy and internal reddening, follows $f_{\nu} \propto \nu^{1/3}$ (Appendix Figure~\ref{sed}) as predicted for accretion disk models (\citealt{1973A&A....24..337S}).

The average host galaxy and AGN fluxes obtained in both 2016 and 2017 campaigns are listed in Table~\ref{table3}. Using linear interpolation of the fluxes obtained from the AGN spectrum, we estimate the monochromatic AGN luminosity $\lambda L_{\rm{\lambda(AGN)}}$ at $5100~$\AA\ to be $L_{\rm{AGN}-2016} = (1.63 \pm 
0.12)\times 10^{44}{\mathrm{erg\ s^{-1}}}$ and $L_{\rm{AGN}-2017} = (1.58 \pm 
0.10)\times 10^{44}{\mathrm{erg\ s^{-1}}}$ for campaigns 2016 and 2017 respectively. To determine the luminosities, we used a distance of 145 Mpc (\citealt{1993AJ....105.1637H}) assuming a standard cosmology with ${H_{0}=73\ \mathrm{km\ s^{-1}\ Mpc^{-1}}}$, $\Omega_{\Lambda}=0.73$ and $\Omega_{m}=0.27$.

\begin{figure*}
  \centering
  \includegraphics[width=\columnwidth]{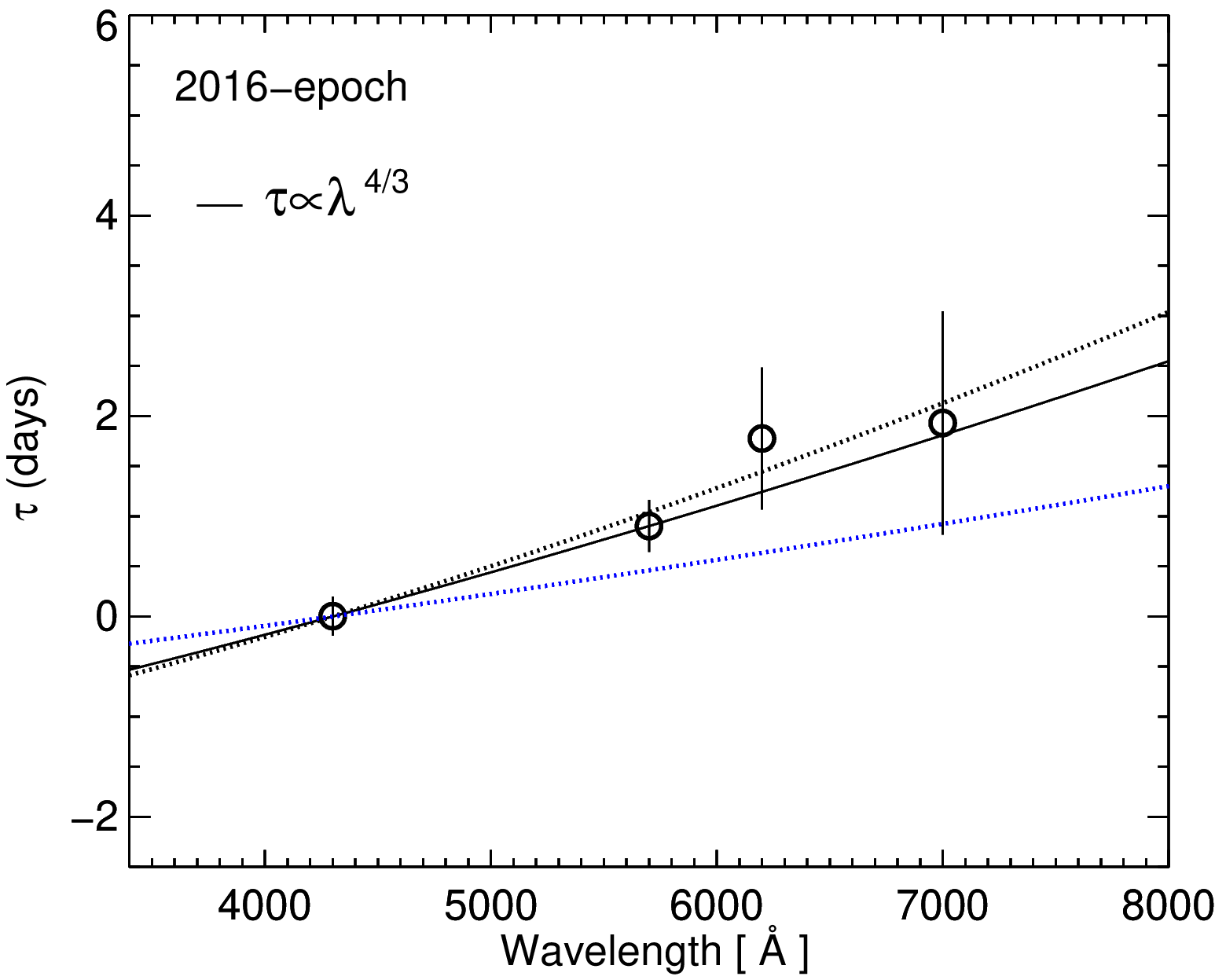}
  \includegraphics[width=\columnwidth]{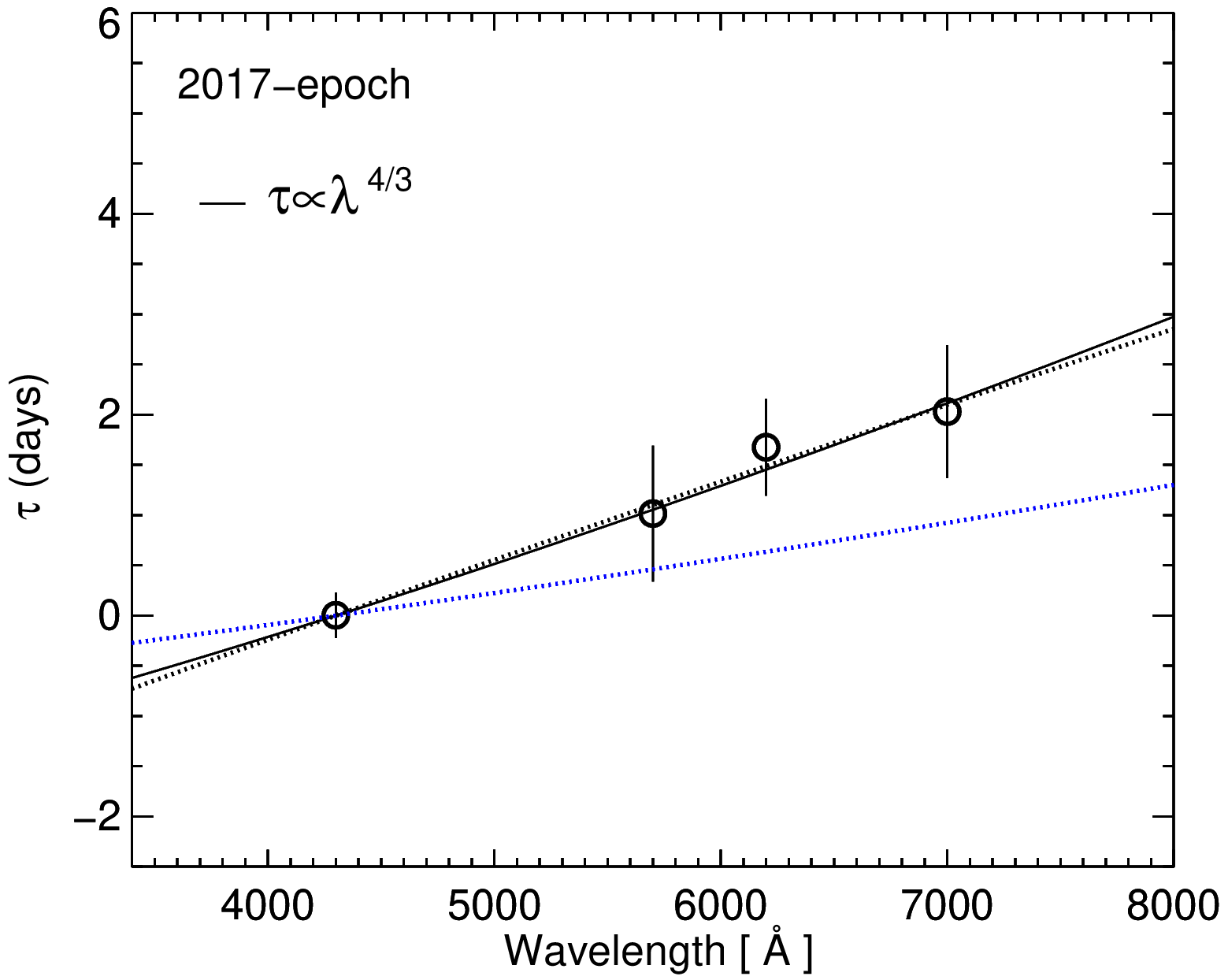}
  \caption{Time delay as a function of wavelength (black open circles) for campaigns 2016 ({\it 
  left}) and 2017 ({\it right}). The expected time delays for an optically thick and geometrically thin accretion disk model are shown by the dotted blue lines. The dotted black line shows the best fit to the observed relation $\tau_{jk} = \alpha (\lambda_{k}^{\beta}-\lambda_{j}^{\beta})$ with $\alpha$ and $\beta$ free parameters. The solid black line is the fit with a fixed theoretically expected index $\beta = 4/3$. The time delays are calculated with respect to the 4300\,\AA\ narrow-band and are corrected by the time dilation factor ($1+z = 1.0344$).}
\label{delayavgres}
\end{figure*}

\section{TIME SERIES ANALYSIS}
In order to robustly estimate the time delays between different continuum bands, we used three different approaches; the traditional interpolated cross-correlation function (ICCF, \citealt{1987ApJS...65....1G}; \citealt{2000ApJ...533..631K}; \citealt{2004ApJ...613..682P}), the discrete correlation function (DCF, \citealt{1988ApJ...333..646E}) including the Z-transformed DCF (\citealt{1997ASSL..218..163A}), and the recently introduced von Neumann statistical estimator (VN; \citealt{2017ApJ...844..146C}); the latter one does not rely on interpolation and binning of the light curves but on the level of randomness of the data. Since the VN estimator is not widely known in RM analysis, we give some comprehensive explanations here.

First we create a combined time series $F$ between the driving $F_{1}$ and time-delayed $F_{2}^{\tau}$ continuum light curves so that $F(t,\tau) = \{(t_{i},f_{i})\}^{N}_{i=1} = F_{1} \cup F_{2}^{\tau}$, with $F_{2}^{\tau} = \{(t_{i}+\tau,f_{i})\}^{N_{2}}_{i=1}$, $f_{i}$ the fluxes measured at times $t_{i}$ for each of the light curves and $N = N1+N2$ correspond to the total number of data points. The VN estimator of the randomness of the combined light curve is defined as the mean-square of successive differences,

\begin{align}
  \mathcal{V_{N}(\tau)} = \frac{1}{N-1} \sum_{i=1}^{N-1} \frac{\left[ F(t_i) -
      F(t_{i+1})\right]^2}{W_{i,i+1}}
\label{equ:vn}
\end{align}
where $W_{i,i+1} = 1/[\sigma_{lc}^2(t_i) +  \sigma_{lc}^2(t_{i+1})]$ is a weighting factor introduced by \cite{1994A&A...286..775P} which takes into account the flux uncertainty ($\sigma_{lc}$) from the light curves (see \cite{2017ApJ...844..146C} for a slightly modified version of this factor). The goal is to find a time delay $\tau_{0}$ from a pre-defined search interval $[\tau_{min},\tau_{max}$] that will minimize the VN estimator so that $V_{N}(\tau_{0}) \equiv min[V_{N}(\tau)]$.

For the three methods we used a common time-delay search interval $[\tau_{min},\tau_{max}$] = $[-10,10]$ days, and we estimated the delays relative to the 4300\,\AA\, narrow-band. For the ICCF, we used the search interval spaced by 0.1 days, while the DCF was evaluated using a bin size of one day which corresponds to the median sampling of the light curves. Since the light curves are very well sampled, the choice of a lower or higher time-bin size does not change the results.
For the ICCF and DCF we estimated the time delay using the centroid $\tau_{cen}$ of the cross-correlation function $R(\tau)$ computed above the correlation level at $R \geq 0.7R_{max}$, except for the pairs 4300/7000 in campaign 2017, where we used $R \geq 0.6R_{max}$ due to the lower correlation found between the bands.

Uncertainties in the time delay were calculated using the flux randomization and random subset selection (FR/RSS) method of \cite{2004ApJ...613..682P} considering the improvements presented by \cite{1999PASP..111.1347W}. From the observed light curves we create 2000 randomly selected subset light curves, each containing 63\% of the original data points due to the non-selection of points according to Poisson probability. The flux value of each data point was randomly altered consistent with its normal-distributed measurement error. We calculated the ICCF and DCF for the 2000 pairs of subset light curves and used the 68\% confidence range to estimate the errors of the centroid. The time delay measurements obtained by various methods are shown in Figure~\ref{delayccf} in the Appendix. Table~\ref{tabledelay} gives the centroid and the central 68\% confidence intervals of the distributions obtained from the FR/RSS method. The time delays obtained with different methods yield consistent results for both 2016/2017 campaigns, although the errors decreased during 2017 campaign due to the higher time sampling of the light curves.

\section{DISCUSSION}

In the following section, we discuss the results in the context of emission from an optically thick and geometrically thin accretion disk.

\subsection{Photometric reverberation mapping of the accretion disk}

According to the standard disk theory of \cite{1973A&A....24..337S}, the energy flux radiated, due to a viscous heating process, from a surface unit of an optically thick and geometrically thin accretion disk is

\begin{align}
  \mathcal{E_{V}}(R) = \frac{3GM\dot M}{8\pi R^3} \left[ 1 - \left(\frac{R_{0}}{R}\right)^{1/2} \right]
\label{equ:energyvis}
\end{align}
where $R$ is the distance away from the innermost radius $R_{0}$ of the disk, $G$ is the gravitational constant, $M$ is the mass of the black hole and $\dot M$ the mass accretion rate of the disk. The boundary of the disk is assumed here to be located at $R_{0}$, also referred as the radius of the innermost stable circular orbit around the black hole and for which only the critical flux of the matter can go under $R_{0}\sim3 R_{g}$ with $R_{g} = 2GM/C^2$ the Schwarzschild radius. Apart from a viscous heating process, the disk photosphere is irradiated by an external UV/X-ray-emitting source with luminosity $L_{*}$. Since the geometry of the emitting source is unknown, a simplistic approximation is made by placing the source at a height $H_{*}$ along the rotational axis of the black hole (see Fig.3 in \citealt{2005ApJ...622..129S}). Considering the albedo $a$ of the disk, the irradiated flux can be expressed as

\begin{align}
  \mathcal{E_{I}}(R) = \frac{L_{*}(1 - a)}{4\pi R^3} H_{*} \cos{\theta}
\label{equ:energyxray}
\end{align}
with $\theta$ the angle between the disk surface normal and the incoming radiation of the emitting source (\citealt{2013peag.book.....N}). The total observed flux from the accretion disk is therefore $\mathcal{E}(R) = \mathcal{E_{V}}(R) + \mathcal{E_{I}}(R)$. If the radius $R$ is much greater than the innermost radius ($R\gg R_{0}$) and assuming that the local emission is described by a perfect blackbody so that $\mathcal{E} = \sigma T^4$, the temperature across the disk is

\begin{align}
  T(R) = \left[ \frac{3GM\dot M}{8\pi R^3\sigma} + \frac{L_{*}(1 - a)}{4\pi R^3\sigma} H_{*} \cos{\theta} \right]^{1/4}
\label{equ:energytot}
\end{align}

The combined temperature profile $T\propto R^{\, -3/4}$ is responsible for the thermal radiation emitted over a range of wavelengths centred at $\lambda_{0} = xhc/kT(R)$, where $x$ is a factor needed in the conversion from $T$ to $\lambda$ for a given radius $R$. Since the factor $x$ depends on the function that is used to describe the radius response to the emitted radiation, we set $x = 2.49$ obtained by adopting a flux-weighted mean radius $\langle R \rangle = {\int_{R_0}^{\infty} B(T(R))R^2\,dR}/{\int_{R_0}^{\infty} B(T(R))R\,dR}$ (\citealt{2016ApJ...821...56F}; \citealt{2017ApJ...840...41E}), with $B(T(R))$ the Planck function, and assuming the temperature profile described in Equation~(\ref{equ:energytot}). The variable radiation from the innermost part of the disk, closer to the black hole, will have the peak of the emission at shorter wavelengths and due to reprocessing effects, the variability is observed with a time delay $\tau$ with respect to the outer and cooler parts of the disk which are traced by longer wavelengths. This effect can be interpreted as the light travel time across the disk so that $\tau = R/c$. In consequence, for two different continuum light curves with central wavelengths at $\lambda_{j}$ and $\lambda_{k}$, the predicted time delay $\tau_{jk}$ between the bands is given by

\begin{align}
  \tau_{jk}= \gamma \left[\lambda_{k}^{4/3} - \lambda_{j}^{4/3} \right] \left[ \frac{3GM\dot M}{8\pi \sigma} + \frac{L_{*}(1 - a)}{4\pi \sigma} H_{*} \cos{\theta} \right]^{1/3}
\label{equ:delayfunction}
\end{align}
with $\lambda_{k} > \lambda_{j}$, and $\gamma = c^{-1}(xk/hc)^{4/3}$. We note that a simplified version of Equation~(\ref{equ:delayfunction}) can be obtained by assuming that the ratio of external to internal heating of the disk ($\kappa = 2L_{*}(1-a)H_{*}/ GM\dot M$) is close to zero, i.e. the contribution of the external UV/X-ray radiation above the disk plane $\mathcal{E_{I}}$ (Equation~\ref{equ:energyxray}) is negligible compared to internal viscous dissipation (see  Equation [5] of \citealt{1998ApJ...500..162C} and Equation [3] of \citealt{2017ApJ...840...41E}).

Figure~\ref{delayavgres} shows the rest-frame average time delay as a function of the central wavelength ($\lambda_{0}/[1+z]$ in Angstroms) obtained for both 2016 and 2017 campaigns. We fit the time delays with the model $\tau_{jk} = \alpha (\lambda_{k}^{\beta}-\lambda_{j}^{\beta})$ with $\alpha$ and $\beta$ as free parameters. The best fit for campaign 2016 is obtained with $\alpha = 2.27\pm0.48$ days and $\beta = 1.48\pm0.61$ and with $\alpha = 2.11\pm0.36$ days and $\beta = 1.21\pm0.42$ for campaign 2017. We then fix $\beta = 4/3$ in order to test the time delay-wavelength relation as predicted for an optically thick and geometrically thin accretion disk model. The best fit is obtained with $\alpha = 2.07\pm0.28$ days for 2016 and $\alpha = 2.16\pm0.19$ for 2017. The measured rest-frame delays can be well-fitted by the standard disk model $\tau \propto \lambda^{4/3}$ in both observing campaigns, albeit with lower uncertainties for both $\alpha$ ($\sim17$\%) and $\beta$ ($\sim35$\%) parameters during 2017 monitoring. This is expected because of the higher time-sampling obtained for the light curves in 2017 ($\sim 0.8$ days) which leads to improved time delay measurements with lower average uncertainties.

\subsection{The accretion disk size of Mrk509}

The observed continuum time delays can be compared with those expected from the standard disk theory for a given black hole mass and mass accretion rate. Assuming a bolometric luminosity correction $L_{\rm Bol} = 10\lambda L_{\lambda}(5100$\AA) (\citealt{2004MNRAS.352.1390M}), a black hole mass of $M = 14.3 \pm 1.2 \times 10^{7} M_{\odot}$ (\citealt{2004ApJ...613..682P}), and a mass to radiation conversion efficiency $\eta = L_{\rm Bol}/\dot M c^2 = 0.10$ (\citealt{2009ApJ...690...20S}), we estimate the mass accretion rate $\dot M_{2016} = 0.29 M_{\odot} yr^{-1}$ and $\dot M_{2017} = 0.28 M_{\odot} yr^{-1}$ for 2016 and 2017 campaigns respectively. The AGN luminosity did not change between the two years, hence the mass accretion rate remained constant. During an XMM-Newton monitoring carried out in 2011 focusing on Mrk509, \cite{2011A&A...534A..39M} estimated a range for the mass accretion rate of $0.24 \leq \dot M \leq 0.34 M_{\odot} yr^{-1}$. Their average value $<\dot M> = 0.29 M_{\odot} yr^{-1}$ is exactly the same as the average value we obtained for 2016 and 2017 campaigns.

\begin{figure}
  \centering
  \includegraphics[width=\columnwidth]{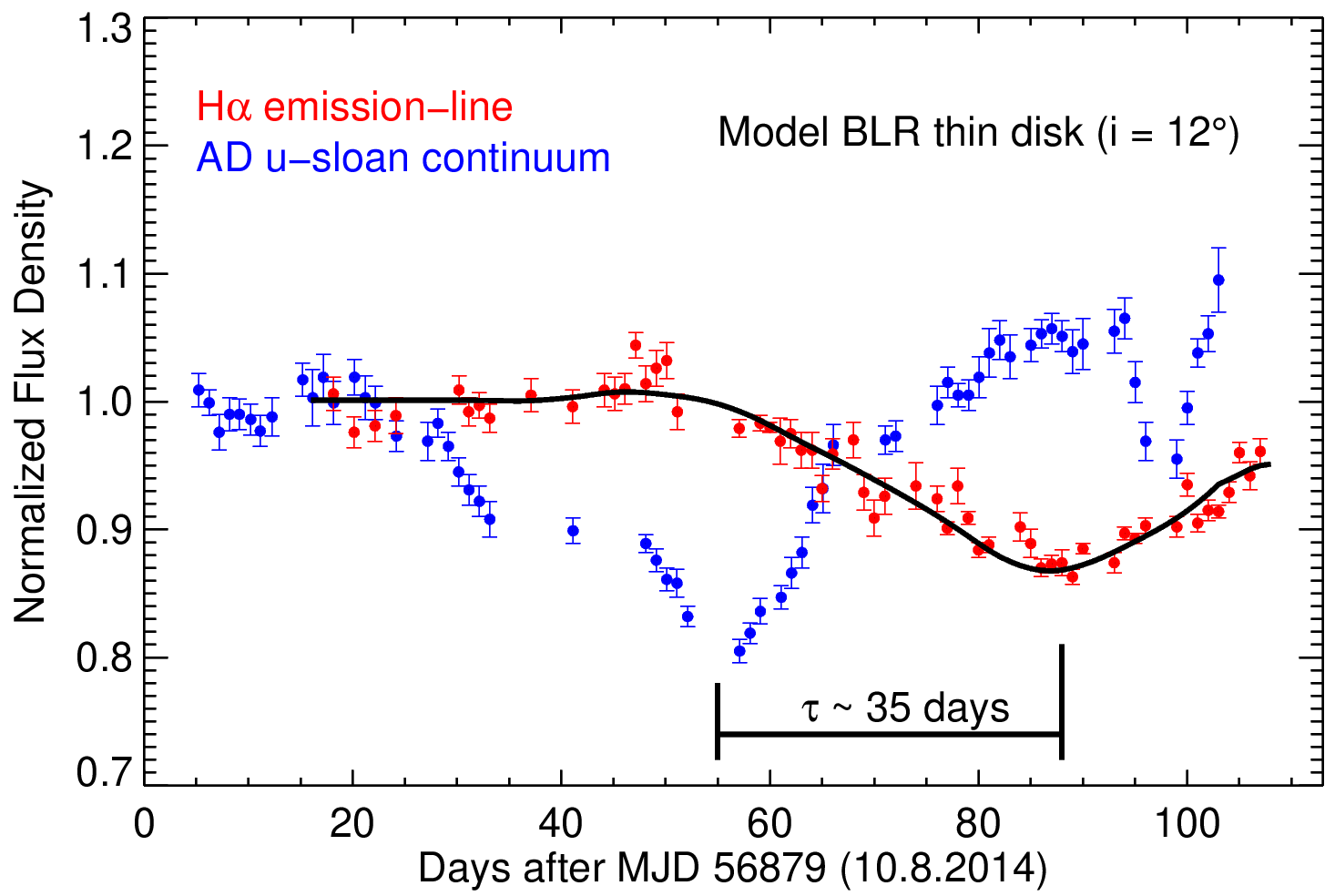}
  \caption{BLR thin-disk model. The blue and red dots show the observed (host-galaxy corrected) UV continuum and H$\alpha$ light curves respectively. A thin-disk BLR model (black solid line) that extends from 32 to 43 light days and has an inclination $i=12^{\circ}$ is able to reproduce the features of the H$\alpha$ light curve.}
\label{blr_geo}
\end{figure}

Given our determination of the mass accretion rate and the black hole mass, we calculate the expected rest-frame time delay with respect to the reference wavelength 4300\,\AA\ (Equation~[\ref{equ:delayfunction}]) to be $\tau = 1.16$ days. Our measured rest-frame delay is a factor of 1.8 larger than the predicted by the standard disk model. An explanation for such a discrepancy could be that the black hole mass of Mrk509 is underestimated. The black hole mass reported by \cite{2004ApJ...613..682P} has been calculated assuming a geometry-scaling factor $f = 5.5$ (\citealt{2004ApJ...615..645O}). The UV continuum and H$\alpha$ emission-line variability observed during 2014 campaign revealed that the echo of the BLR has a mean lag of $\sim35$ days (Blex et al. in prep). In order to constrain the $f$ value from the BLR data, we modelled the H$\alpha$ light curve assuming Keplerian orbits, thin/thick disks and spherical BLR geometries. The modelling follows that of \cite{2014A&A...568A..36P}. The convolution of the UV continuum light curve with a thin-disk BLR model at inclination $i=12^{\circ}$ provides an acceptable fit to the observed H$\alpha$ data (Figure~\ref{blr_geo}). If the BLR of Mrk509 has a nearly face-on disk-like BLR geometry, the geometry-scaling factor is $f=\frac{2 \cdot \ln2}{\sin^2{i}} = 32$. This is about 6 times larger than the commonly used average value obtained by \cite{2004ApJ...615..645O}, and which assumes that AGN and quiescent galaxies follow the same $M_{BH}-\sigma_{*}$ relationship. Therefore, if we adopt $f = 32$ it results in a black hole mass of $M \sim 8 \times 10^{8} M_{\odot}$. In the next section, we describe the implications of the thin-disk BLR geometry for the observed time delays.

\begin{figure*}
  \centering
  \includegraphics[width=2.0\columnwidth]{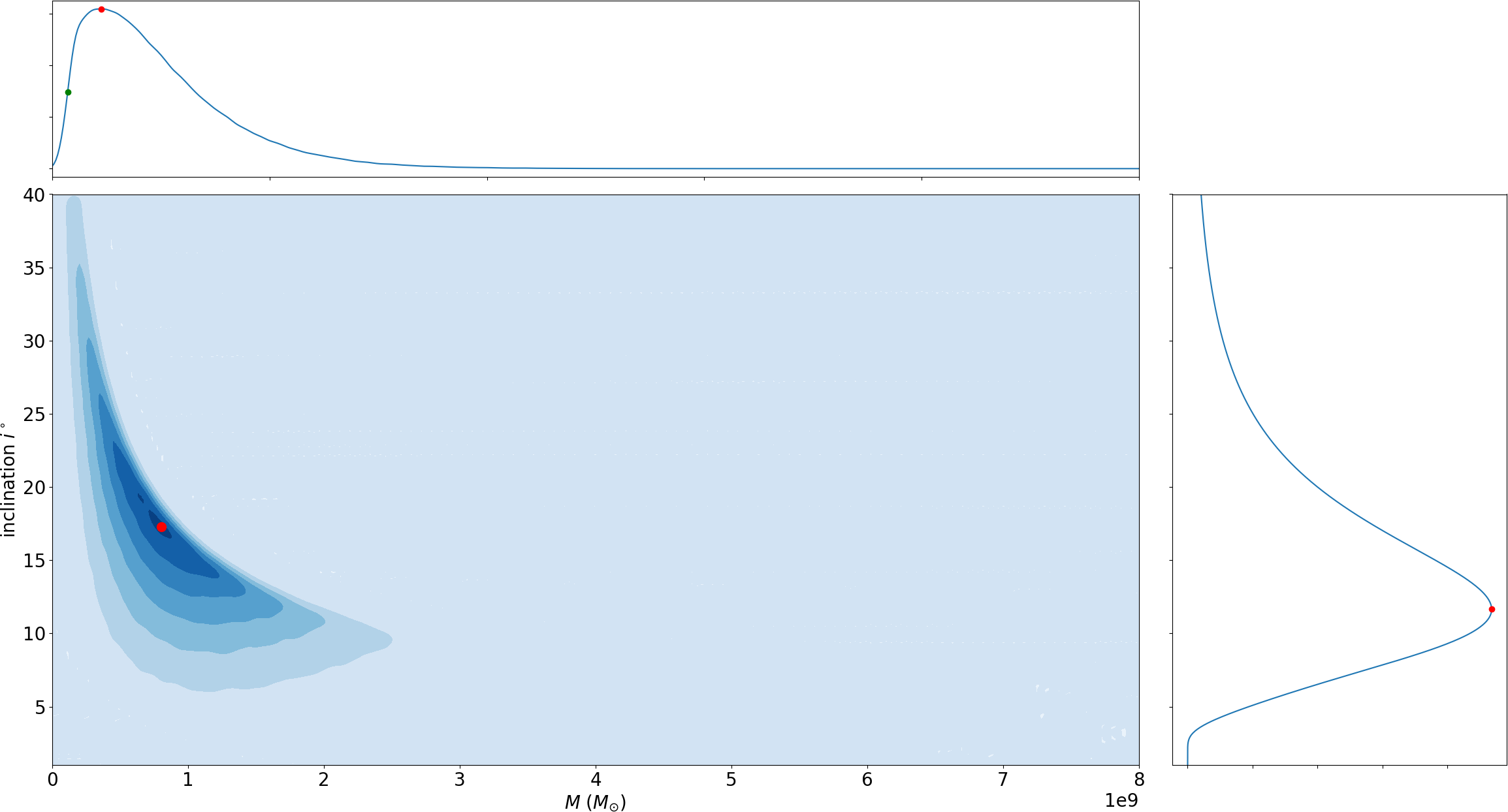}
  \caption{Posterior distribution for mass-inclination  $p(M, i|\mathcal{D})$. The plot shows how much the data support $M, i$ pairs as candidate solutions; the stronger the hue, the more likely the pair. The marginal $p(i|\mathcal{D})$ (top) shows the distribution of $i$ after accounting of the effect of all other model parameters. Similarly, we show the marginal $p(M|\mathcal{D})$ at the right.
  While the most likely pair is the pair $M = 7.82\times 10^{8} M_{\odot}, i = 17.49^{\circ}$, shown as a red dot in the $p(M, i|\mathcal{D})$ diagram, the posterior reveals that other pairs are also probable. We note that in general, the mode of the join distribution does not coincide with the mode of the marginals.
  An indication of the range of probable distinct values for $i$ and $M$ can be read in the marginal distributions, with the most likely value also marked as a red dot for each case. Additionally, we plot as a green dot the previous estimate for the mass $M = 14.3 \pm 1.2 \times 10^{7} M_{\odot}$ which under our probabilistic analysis now appears as a less likely estimate. }
\label{fig:jointden}
\end{figure*}

\subsubsection{Accretion disk probabilistic modelling}

We carry out a probabilistic analysis in order to explore the set of likely solutions for the thin AD model $\tau \propto \lambda^{4/3} (M\dot M)^{1/3}$
whose geometry constrains the black hole mass $M$ (Figure \ref{blr_geo}).

We set the velocity dispersion of the H$\alpha$ emission-line to $V_{\rm H\alpha} = 1730\pm 400\, km/s$ (Blex et al. in prep), and the mass accretion rate to $\dot M = L_{\rm Bol}/\eta c^{2}$, assuming a bolometric correction $L_{\rm Bol} = 10\lambda L_{\lambda}(5100$\AA). Since the radiative efficiency $\eta$ depends on the spin of the black hole, we set $\eta = 0.1$ which assumes that the black hole is co-rotating with the disk\footnote{The radiative efficiency $\eta$ can vary between 0.038 and 0.42 depending on the spin of the black hole (\citealt{2011ApJ...728...98D}). A value of $\eta \geq 0.1$ is commonly used for co-rotating disks, smaller or larger values will underestimate/overestimate the mass accretion rate.}. Under the previous assumptions, we define the time-delay function $\phi \propto \lambda^{4/3}\ (\sin^{-2}(i)\ R_{\rm BLR}\ V^{2}_{\rm H\alpha}\ L_{5100\angstrom})^{1/3}$, where the free model parameters are  $\bd{\theta}=(i,R_{\rm BLR}, L_{5100\angstrom})$. The observed data  $\mathcal{D} = \{(\tau_n, \sigma_n, \lambda_n)\}_{n=1}^N$  are taken from the 2017 campaign (Table \ref{tabledelay}). The  goal of this analysis is to infer the posterior distribution $p(\bd{\theta}|\mathcal{D})$ of the thin AD model parameters. This distribution will help us to examine whether the observed data support our BLR geometry assumption which attempts to reconcile the apparent discrepancy between observed and theoretical time delays.

We formulate the following probabilistic model via the likelihood function $p(\mathcal{D}|\bd{\theta})$:

\begin{align}
p(\mathcal{D}|\bd{\theta}) &= 
\prod_{n=1}^N p(\tau_n| \lambda_n, i, R_{\rm BLR}, L_{5100\angstrom},  \sigma_n)  \nonumber \\
&= \prod_{n=1}^N \mathcal{N}(\tau_n| \phi_{rel}(\lambda_n; i, R_{\rm BLR}, L_{5100\angstrom}), \sigma_n) 
\label{eq:likelihood}
\end{align}
where $\mathcal{N}(x|a,b)$ is the normal distribution with mean and standard deviation $a$ and $b$ respectively. The mean of the normal distributions in Equation~[\ref{eq:likelihood}] is given by the delay-wavelength relation relative to $4300\angstrom$:

\begin{align}
    \phi_{rel}(\lambda_n; i, R_{\rm BLR}, L_{5100\angstrom}) & = 
    \phi(\lambda_n; i, R_{\rm BLR}, L_{5100\angstrom}) \nonumber \\
    & \ - \phi(4300\angstrom; i, R_{\rm BLR}, L_{5100\angstrom})
\end{align}

We complete the probabilistic formulation by imposing prior distributions on the model parameters
$p(\bd{\theta})=p(i)\ p(R_{\rm BLR})\ p(L_{5100\angstrom})$ with $p(i)=\mathcal{U}(i|0.0^\circ, 
40.0^\circ)$ and $p(R_{\rm BLR})=\mathcal{U}(R_{\rm BLR}|10.0, 100.0\ {\rm days})$ where 
$\mathcal{U}(x|a,b)$ is the continuous uniform distribution with support $[a,b]$. Regarding 
luminosity, we impose the normal prior $\mbox{$p(L_{5100\angstrom})=\mathcal{N}(L_{5100\angstrom} | 1.5761068\times 10^{44}, 0.10\times 10^{44}\ {\mathrm{erg\ s^{-1}}})$}$ informed by the luminosity obtained in the 2017 campaign (Section 2.2).

The joint posterior of the model parameters is given by Bayes' theorem\footnote{In order to compute the posterior, we discretize the support of each physical parameter on a grid of $400$ number of equidistant grid points. This turns the integrals into easily computable sums.
While this numerical approach is feasible for our case of three model parameters, it is impractical for more parameters.}:
\begin{align}
    p(\bd{\theta} | \mathcal{D}) =
    \frac{p(\mathcal{D}|\bd{\theta}) p(\bd{\theta})}{ \int p(\mathcal{D}|\bd{\theta})p(\bd{\theta}) \bd{d\theta} }    
\end{align}
Our aim is to verify whether the hypothesis of a BLR thin-disk geometry with a low inclination $i$ and an upwards revised estimate for the mass $M$, is consistent with the observed data. Hence, the specific quantity we seek to infer is the joint distribution $p(M, i|\mathcal{D})$.
This can be computed numerically by first drawing  a large number of samples from the posterior
\begin{align}
i, R_{\rm BLR}, L_{5100\angstrom} \sim 
p(i, R_{\rm BLR}, L_{5100\angstrom} | \mathcal{D}) = p(\bd{\theta}|\mathcal{D})
\end{align}

\begin{figure}
  \centering
  \includegraphics[width=\columnwidth]{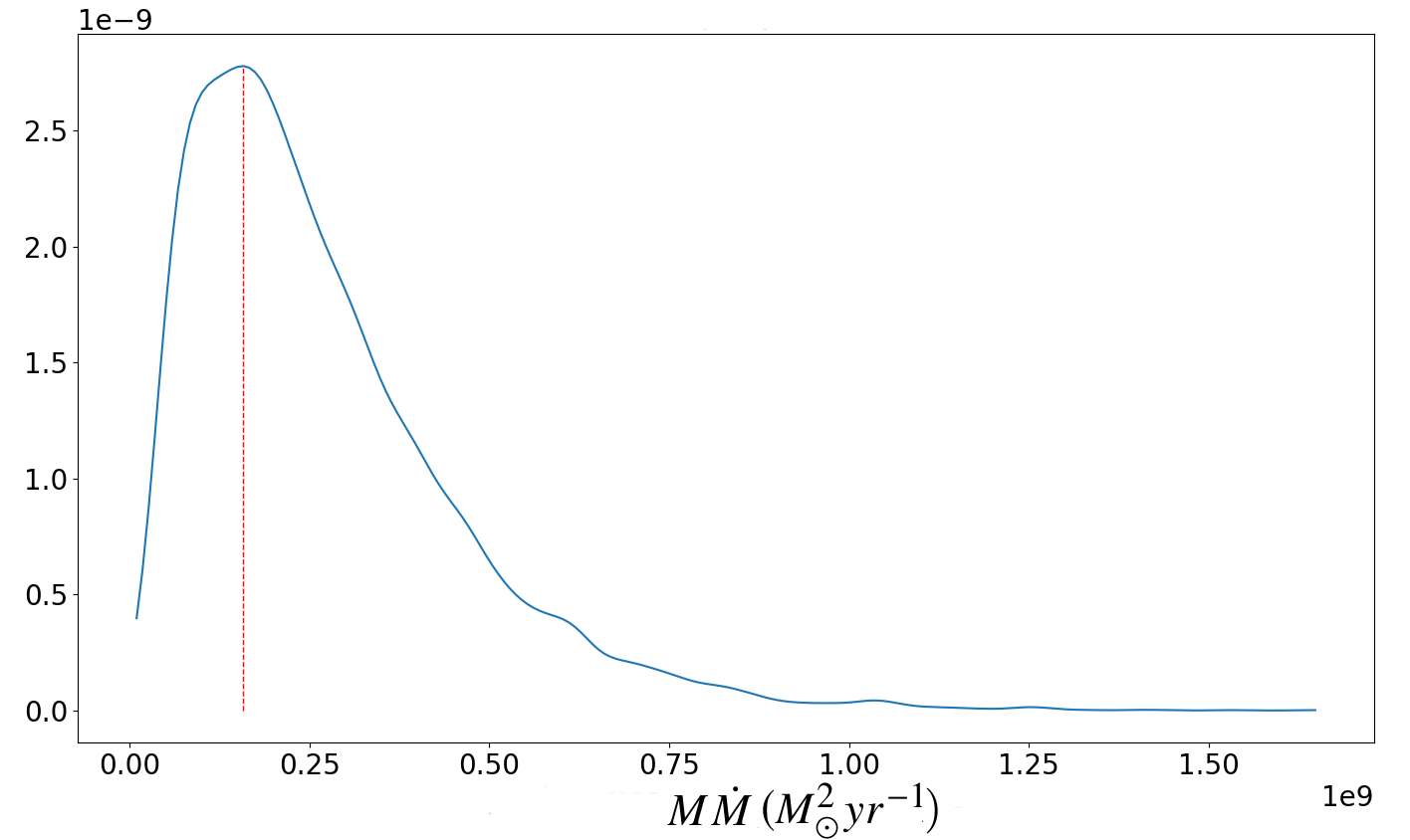}
  \caption{Probability distribution for $M\dot M$. The vertical dotted line marks the mode $M \dot M = 1.58\times 10^{8} M_{\odot}^{2} yr^{-1}$ of the distribution.}
\label{fig:mass_product}
\end{figure}

\begin{figure*}
  \centering
  \includegraphics[width=15cm,clip=true]{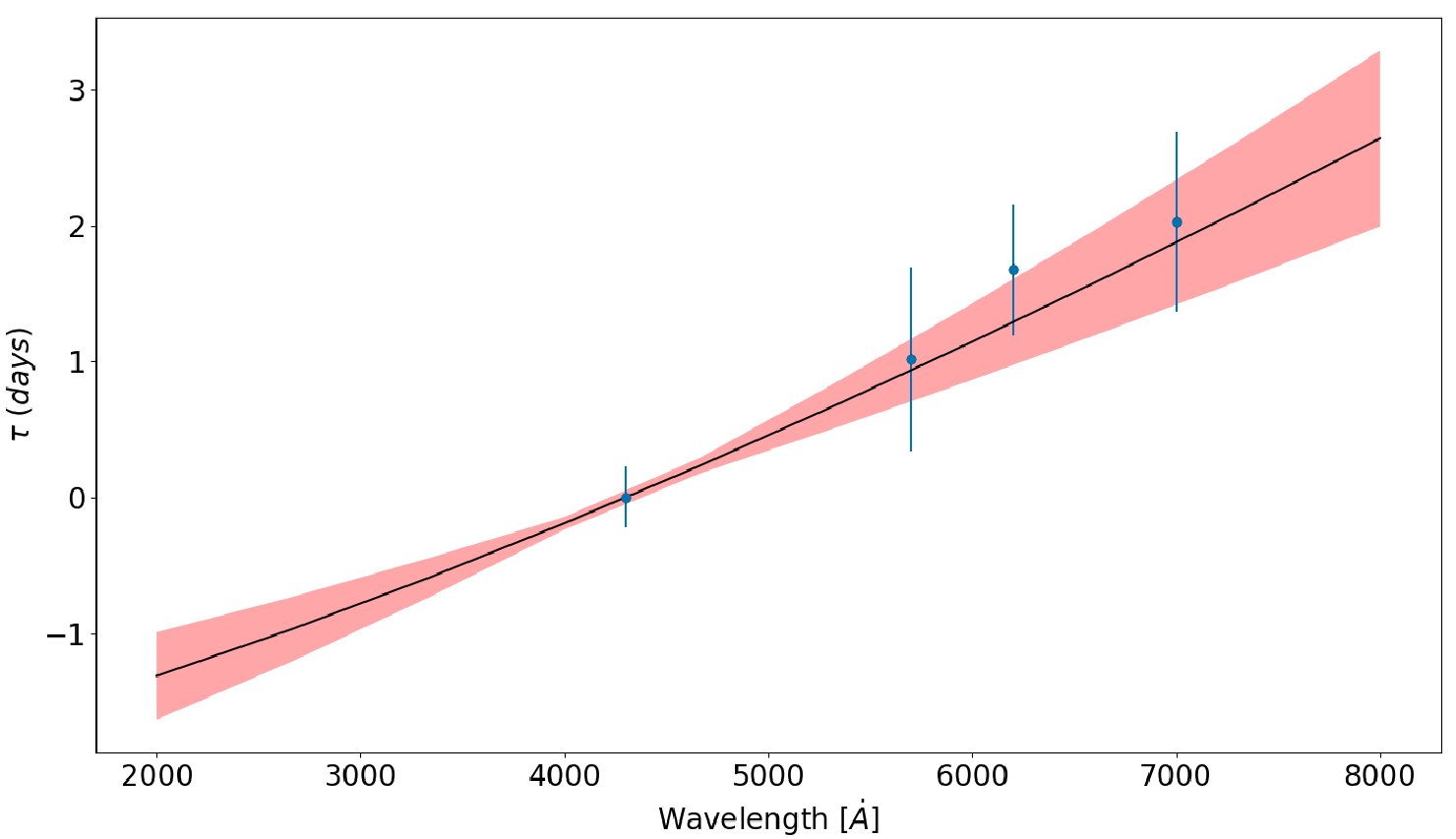}
  \caption{Predictive distribution of AD models along with the data from the 2017 campaign. The black line is the mean of this distribution (i.e. the mean prediction), while the red shaded area corresponds to $\pm1$ standard deviation from the mean.}
\label{fig:predictive}
\end{figure*}

Samples for the mass parameter $M$ are indirectly obtained from the drawn $R_{\rm BLR}$, $i$ samples. Having obtained a large number of samples, we then use  kernel density estimation to estimate $p(M, i|\mathcal{D})$ from the drawn samples. 
Figure~\ref{fig:jointden} shows the estimated posterior probability $p(M, i|\mathcal{D})$ along with the marginals $p(i|\mathcal{D})$ and $p(M|\mathcal{D})$.  It is evident that the black hole mass obtained with a geometry-scaling factor $f = 5.5$ ($M = 14.3 \pm 1.2 \times 10^{7} M_{\odot}$) is a less likely estimate. The marginal $p(i|\mathcal{D})$ reveals that likely inclinations are roughly in the range $10^{\circ} \leq i \leq 30^{\circ}$; $p(M|\mathcal{D})$ reveals that black hole masses are roughly in the range $0.2 \leq M \leq 1.5 \times 10^{9} M_{\odot}$. The mode of the marginal distribution for the inclination and black hole mass is $i = 11.65^{\circ}$ and $M = 4.77\times 10^{8} M_{\odot}$ respectively. These values are in good agreement with the point estimates derived from the BLR modelling (Figure~\ref{blr_geo}). 

We compute also the probability distribution $p(M\dot M)$ (Figure~\ref{fig:mass_product}). The most likely value for the product is $M \dot M = 2.2\times 10^{8} M_{\odot}^{2} yr^{-1}$. If we consider the mode obtained from the marginal distribution for the black hole mass, the accretion rate is $\dot M = 0.33 M_{\odot} yr^{-1}$, which is consistent with the average value $<\dot M> = 0.29 M_{\odot} yr^{-1}$ obtained for 2016 and 2017 campaigns. Figure~\ref{fig:predictive} plots the posterior predictive distribution for the AD models 
$p(\tau|\lambda) = \int p(\tau|\lambda; \bd{\theta}) p(\bd{\theta}|\mathcal{D}) \bd{d\theta}$ as supported by the posterior $p(\bd{\theta}|\mathcal{D})$. In other words, the plot reveals the distribution of the AD model predictions as weighted by the posterior $p(\bd{\theta}|\mathcal{D})$. The new estimate of $M$ and $\dot M$, obtained for a thin-disk BLR at inclination $i = 12^{\circ}$, increase the accretion disk size to $\sim 2$ days\footnote{It is clear from Equation~(\ref{equ:delayfunction}) that not only a larger value for the black hole mass will result in a larger disk size, for instance, if we assume that all the emission at a certain wavelength comes from an annulus of radius $R$ at a temperature given by Wien's law, the factor $x$ becomes twice as large (4.97), increasing the $\gamma$ factor in Equation~(\ref{equ:delayfunction}) and scaling the disk size by a factor $\sim2.5$. However, the use of a flux-weighted radius is a more realistic assumption since it assumes that the temperature profile of the disk follows $T\propto R^{-3/4}$ as predicted by the standard disk theory of \cite{1973A&A....24..337S}.}, hence consistent with our observations.

Recent continuum reverberation mapping studies have also found accretion disk sizes which are a factor of $\sim 2-3$ larger than predicted by the standard thin disk model (NGC5548; \citealt{2015ApJ...806..129E}, and \citealt{2018ApJ...854..107F} for the Seyfert-1 galaxies NGC2617 and MCG+08-11-011). These previous studies have been carried out using broad-band filters which are contaminated by emission from the BLR and therefore they might bias the time delays to larger values. As shown by \cite{2015ApJ...806..129E} for NGC5548, the observed U-band time-delay of 1.35 days versus a predicted delay of 0.85 days can be explained by accounting for Balmer diffuse continuum emission from the BLR, hence the advantage of using narrow-band filters which are less affected by BLR emission. Another interesting example is NGC2617 ($z = 0.014)$ for which \cite{2018ApJ...854..107F} found a disk size about 2.1 larger than predicted. NGC2617 is about 40 times less luminous than Mrk509 and the authors were able to account for the systematic difference by increasing the product $M\dot M$, although they did not account for a larger black hole mass due to BLR geometry effects.

\cite{2017MNRAS.467..226G} proposed that the larger accretion disk sizes found by previous RM and microlensing studies can be reconciled with the standard disk theory (\citealt{1973A&A....24..337S}) after correcting for AGN internal extinction. Accounting for AGN reddening can increase the optical $L_{5100\angstrom}$ luminosities up to a factor of 4. However, as shown in Section 2.2, Mrk509 has a reddening consistent with $E(B-V) = 0$ and therefore nuclear extinction does not explain the observed larger disk size.

Using the same filter configuration than for Mrk509, \cite{2019NatAs...3..251C} found significantly larger continuum time delays for the Seyfert-1 galaxy Mrk279. The delays observed in Mrk279 followed a supra-linear steep rise with wavelength, implying a different temperature profile than the predicted by the standard thin-disk theory. From reported lags of 14 AGN, \cite{2007ASPC..373..596G} observed a similar steep rise in $\tau$ at long wavelengths. He attributed this to contamination by light being reprocessed from further away. For Mrk279 and the objects considered by \cite{2007ASPC..373..596G}, the steep rise observed in the time-delay with wavelength is a consequence of emission of a farther away, under-appreciated, non-disk component that significantly contributes to the flux at longer wavelengths. Through the use of photo-ionization modelling, \cite{2019NatAs...3..251C} identify this component as high-density, photo-ionized material that has been uplifted from the outer accretion disk, likely due to radiation-pressure force on dust. This supra-linear steep rise with wavelength found in Mrk279 has not been found in Mrk509 and therefore the observed uplifted material from the accretion disk is particularly related to the source.

\section{Conclusions}

We have performed a two-year photometric reverberation mapping monitoring campaign in order to study the optical continuum emission from the nucleus of the Seyfert-1 galaxy Mrk509. The main results are:

\begin{enumerate}

\item We have detected inter-band continuum time delays in two different epochs 2016 and 2017 by using a novel narrow-band imaging experimental design which mitigates the emission line and pseudo-contamination of the signal from the BLR. The results are remarkably consistent between both photometric campaigns, although the time-delay measurements have been improved significantly in 2017 as a consequence of the higher (sub-day) time sampling obtained for the light curves. The measurements are also consistent with the fact that the average bolometric luminosity remained constant during both observing campaigns. These results confirm that time resolution is a crucial factor in order to measure AGN continuum time delays with an accuracy needed to constrain theoretical models of the accretion disk, therefore future coordinated ground-based optical observing campaigns are of vital importance.

\item the time delays increase with wavelength according to the relation $\tau \propto \lambda^{4/3}$ predicted for an optically thick and geometrically thin accretion disk. However, the inferred disk size is larger by a factor of 1.8 than predictions based on the standard thin-disk theory.

\item the larger disk size found in Mrk509 can be explained if the black hole mass is a factor of 3.3 larger than the current value obtained through reverberation mapping of the BLR. This is supported with a probabilistic modelling of the continuum time delays that assumes a BLR with a thin-disk like geometry, and independently corroborated by the direct modeling of the BLR observations. The BLR small inclination $i\sim 12^{\circ}$ leads to a geometry-scaling factor $f$ that is $\sim6$ times larger than the commonly used average value ($f=5.5$). The internal extinction and bolometric luminosity corrections plays a minor role in the particular case of Mrk509.

\item the accretion disk probabilistic modelling of the continuum time delays can be used directly to infer the black hole masses without explicitly accounting for the BLR geometry scaling factor. In that way, the resulting black hole masses can be compared with values obtained by the direct modeling of the BLR emission-line light curves. Since this is clearly a model dependent analysis, it can only be applied if the data are not biased by the effect of inclination or external contamination (e.g. AGN internal reddening, BLR line and diffuse continuum emission), highlighting the importance of using specific designed narrow-band filters. Future applications of this approach on larger data sets will provide more constraints on specific models of the accretion disk.

\end{enumerate}

Although in some cases the black hole masses might not fully reconcile the theory with observations, their uncertainties due to the unknown geometry of the BLR are still important quantities that need to be improved. Microlensing studies have reported larger accretion disk sizes for luminous distant quasars, however, high redshift-quasars have a factor of 10 higher black hole masses than quiescent galaxies, hence a scaling factor $f$ obtained from the $M_{BH}-\sigma_{*}$ relationship may not be valid in general. Whether a BLR with a thin-disk like structure holds for all Seyfert-1 galaxies, the determination of the $f$-factor is crucial in order to constrain the significance of the discrepancies between observations with the standard accretion disk theory. In that context, coordinated ground-based optical monitoring campaigns of the accretion disk and the BLR are necessary to increase the observing time-sampling needed to study specific accretion disk models and to further decrease the biases in the time-delay measurements.

\section*{Acknowledgements}

We are grateful to D. Maoz and D. Chelouche for allowing the use of the C18 telescope in Israel, and S. Kaspi for providing technical support with C18 telescope operation. We thank M. Murphy for providing technical support with the telescope operations at the Cerro Armazones observatory in Chile. This research has been partly supported by grants 950/15 from the Israeli Science Foundation (ISF) and 3555/14-1 from the Deutsche Forschungsgemeinschaft (DFG). We also acknowledge support from the IdP II 2015 0002 64 and DIR/WK/2018/09 grants of the Polish Ministry of Science and Higher Education. This work is based on observations collected at the Wise Observatory with the C18 telescope. The C18 telescope and most of its equipment were acquired with a grant from the Israel Space Agency (ISA) to operate a Near-Earth Asteroid Knowledge Center at Tel Aviv University. Authors N.G and K.L.P gratefully acknowledge the generous and invaluable support of the Klaus Tschira Foundation. This work was supported by the Nordrhein-Westf{\"a}lische Akademie der Wissenschaften und der K{\"u}nste, funded by the Federal State Nordrhein-Westfalen and the Federal Republic of Germany. This research has made use of the NASA/IPAC Extragalactic Database (NED) which is operated by the Jet Propulsion Laboratory, California Institute of Technology, under con- tract with the National Aeronautics and Space Administration. This research has made use of the SIMBAD database, operated at CDS, Strasbourg, France. We thank our referee Martin Gaskell for his constructive comments and careful review of the manuscript.





\bibliographystyle{mnras}
\bibliography{mnras_fpozo}




\appendix

\section{.}

\begin{figure*}
  \centering
  \includegraphics[width=18cm,clip=true]{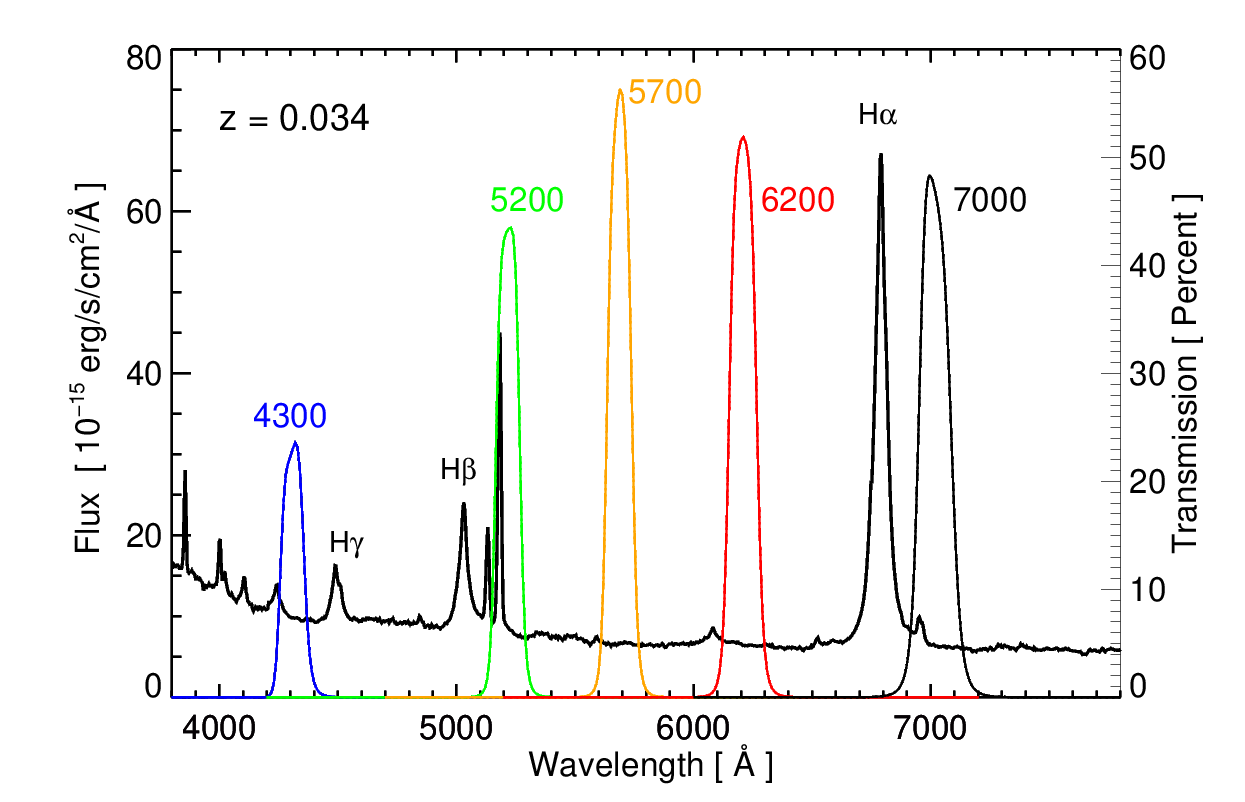}
  \caption{AGN Watch spectrum of Mrk509. The effective transmission of the narrow-band filters used in the monitoring are overplotted in colored lines. The filters curves have been folded with the quantum efficiency of the STL-6303 CCD camera.}
  \label{fig:filters1}
\end{figure*}

\begin{figure*}
\includegraphics[width=0.6\columnwidth]{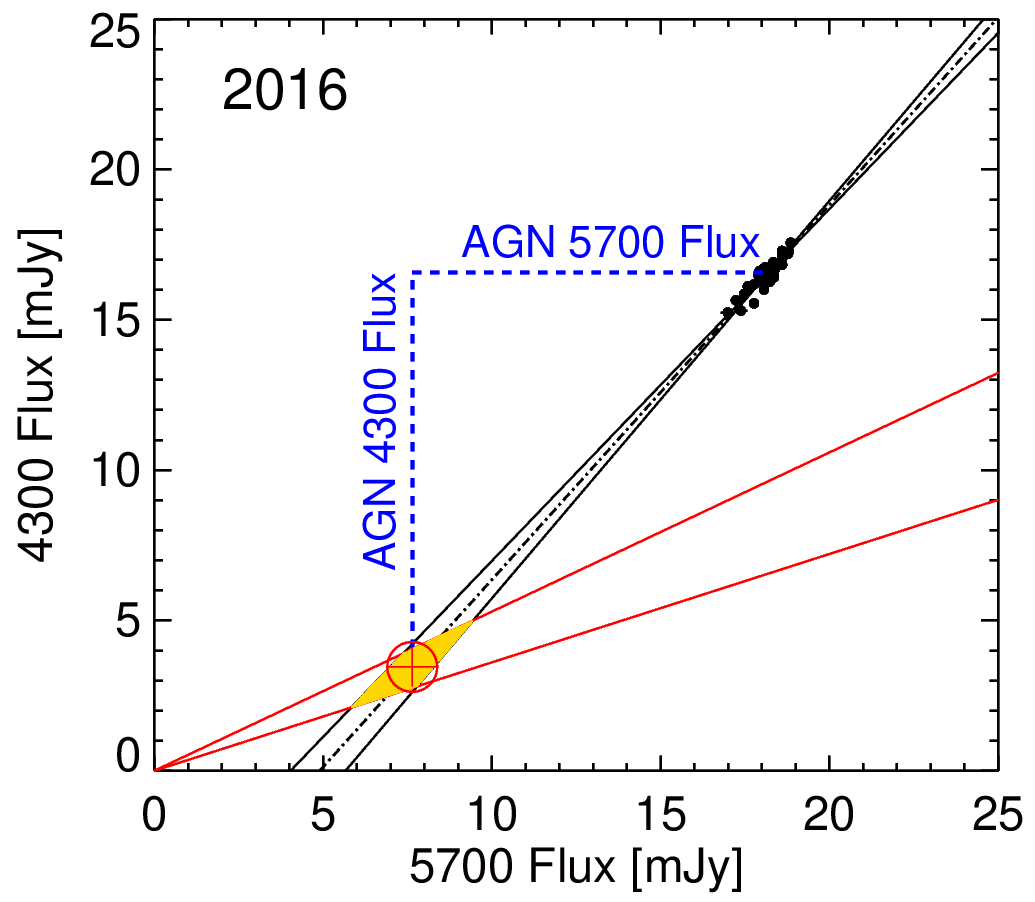} 
\includegraphics[width=0.6\columnwidth]{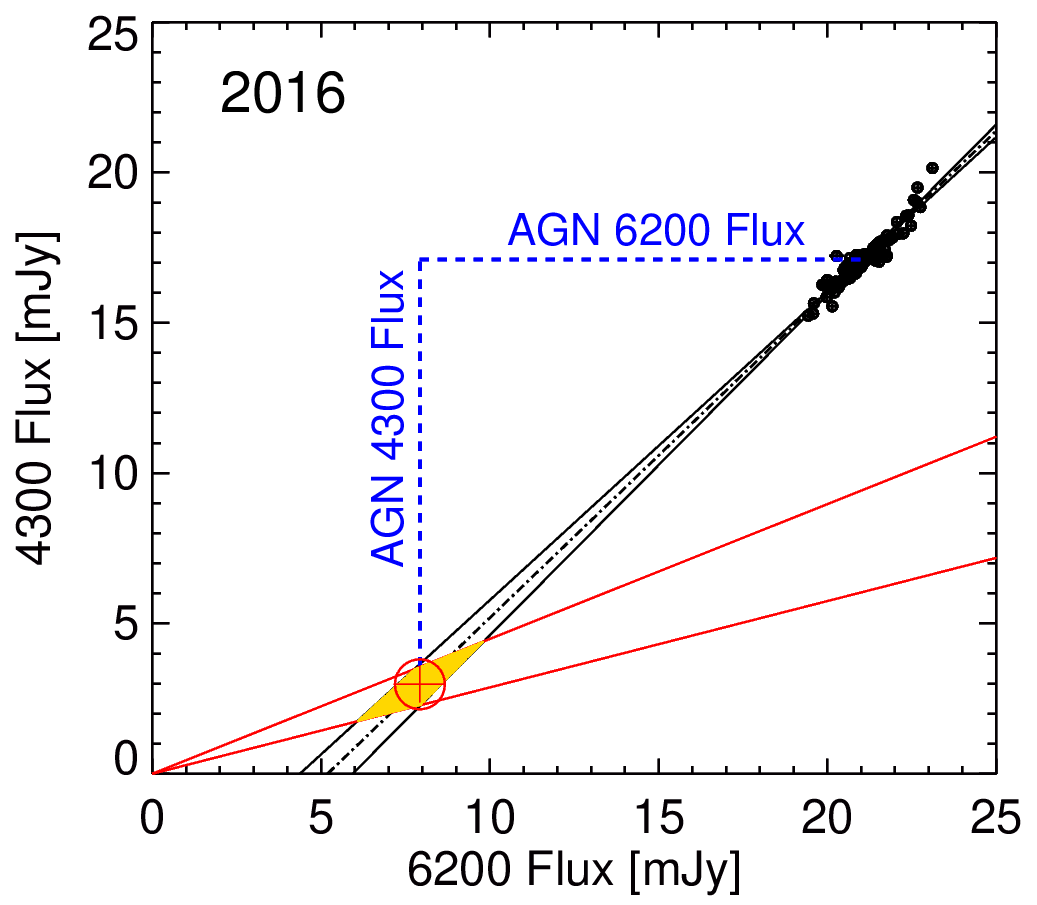} 
\includegraphics[width=0.6\columnwidth]{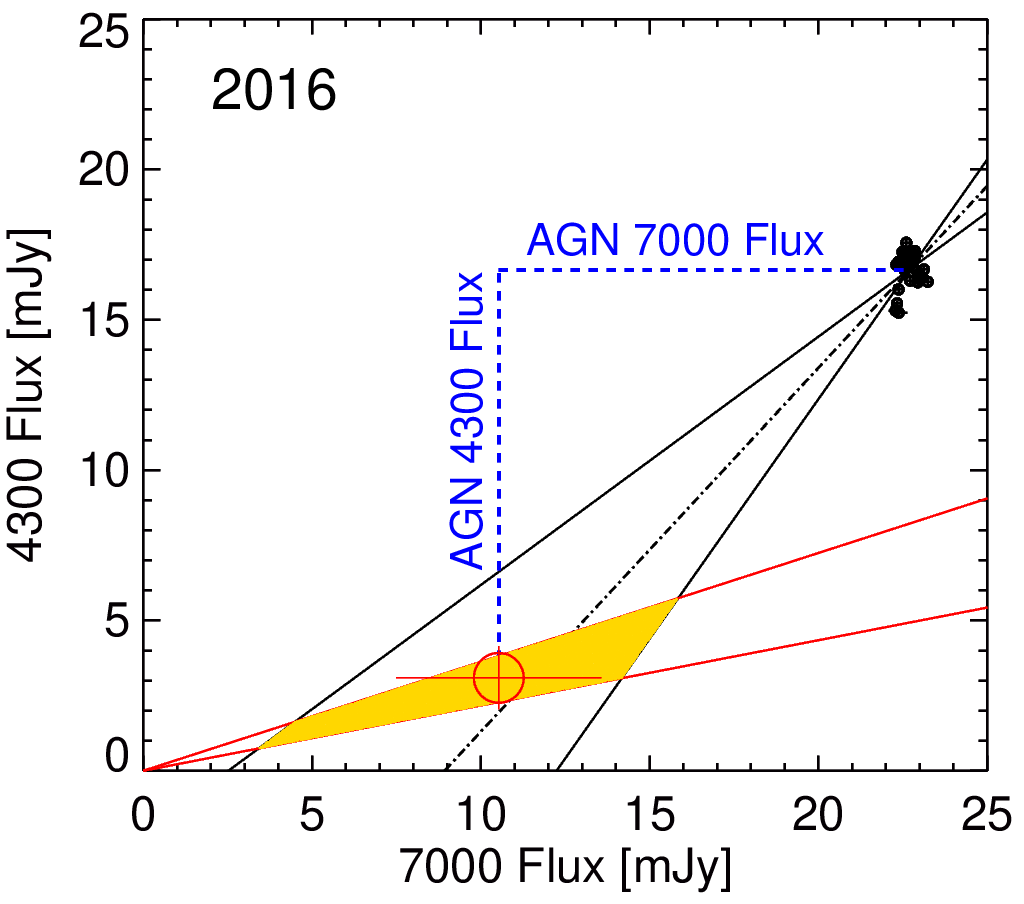}
\includegraphics[width=0.6\columnwidth]{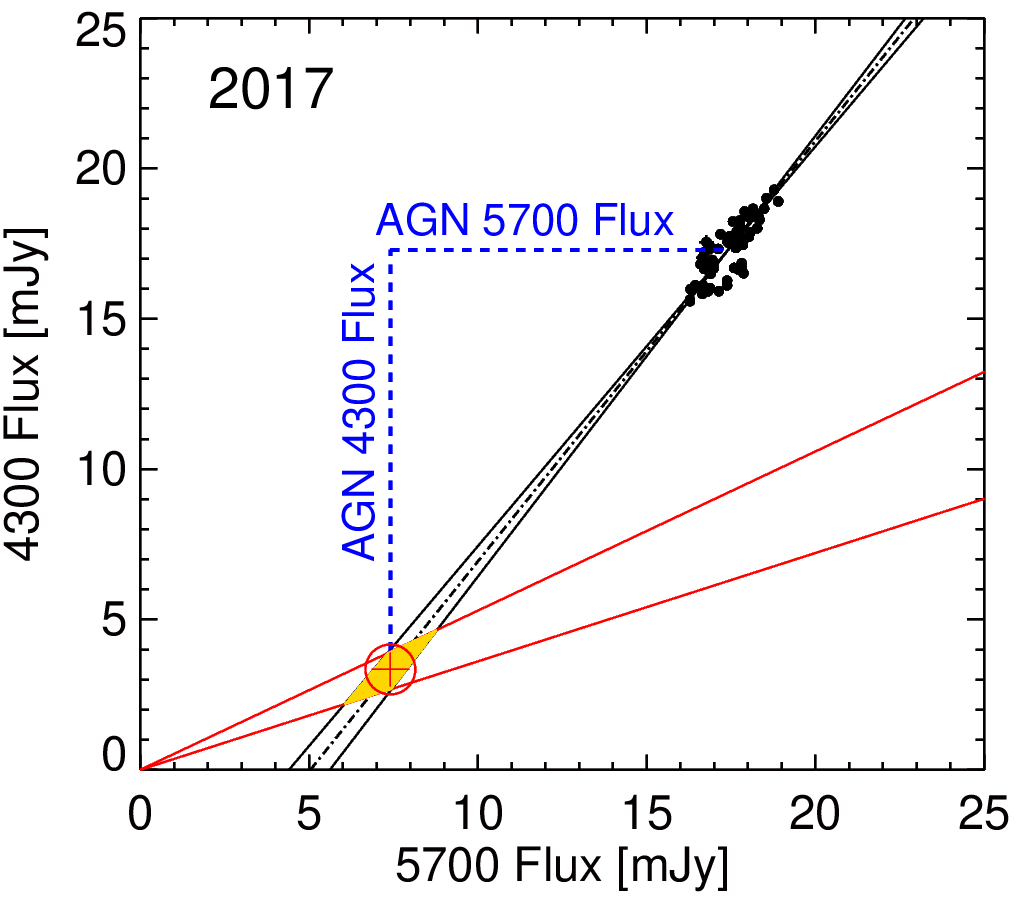} 
\includegraphics[width=0.6\columnwidth]{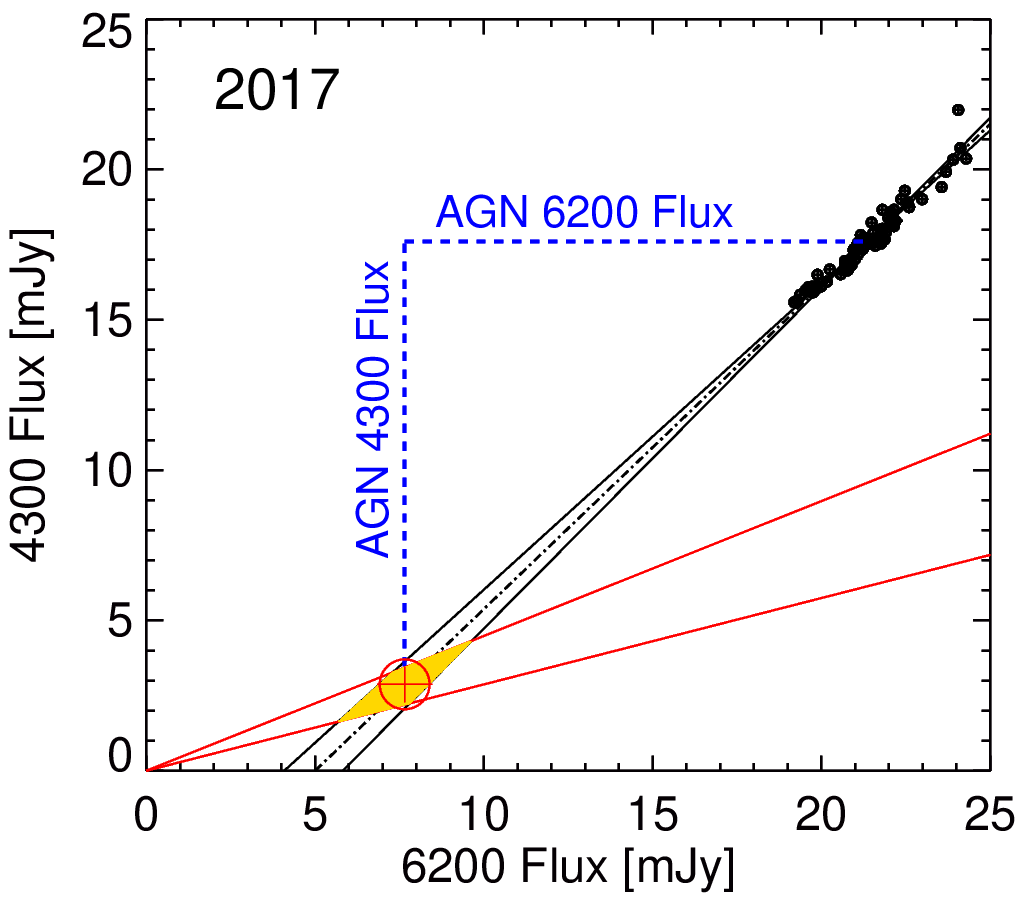} 
\includegraphics[width=0.6\columnwidth]{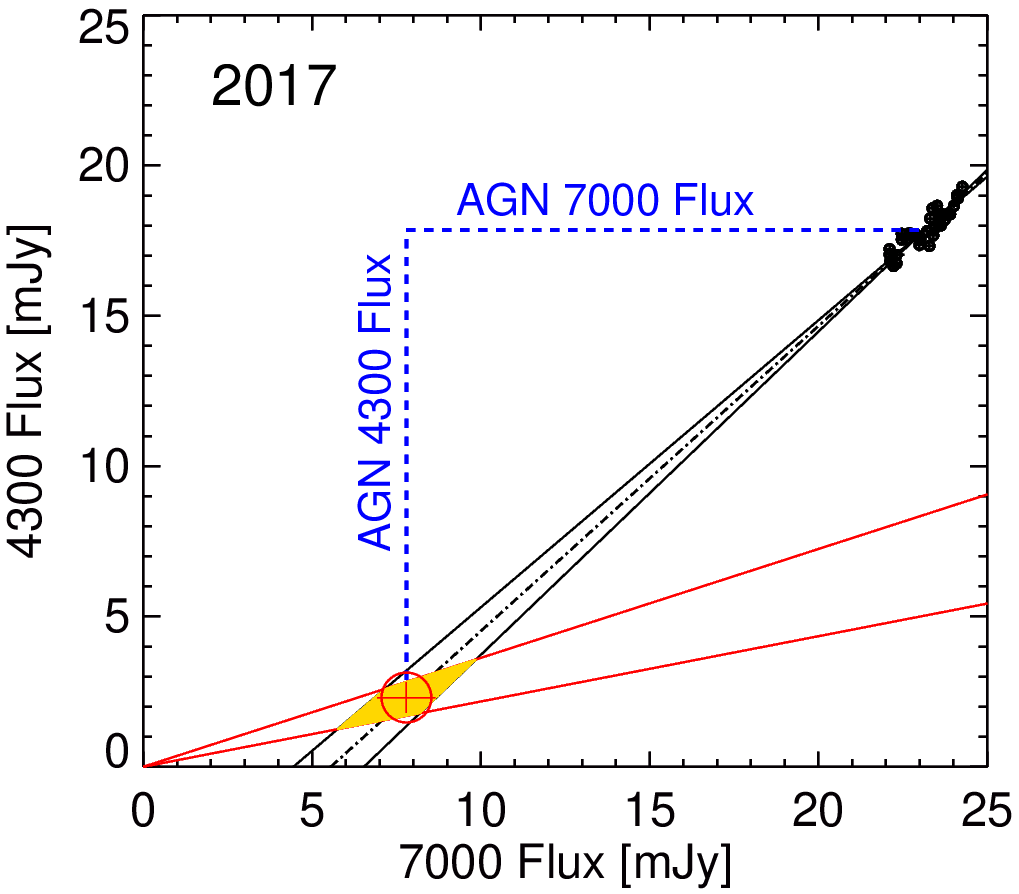}
\caption{Flux variation gradient diagrams for 2016 (top) and 2017 (bottom) campaigns. The solid black lines delineate the ordinary 
least square bisector regression model yielding the range of the AGN slope. The red solid lines indicate the range of host slopes 
obtained from \citealt{2010ApJ...711..461S} for a sample of 11 nearby Seyfert-1 galaxies. The intersection between the host galaxy and AGN slope (yellow area) gives the host galaxy flux in both bands (red circle with cross). The dash-dotted blue lines depict the 
range of the host subtracted AGN flux in both bands.}
\label{fvg}
\end{figure*}

\begin{figure*}
  \centering
  \includegraphics[angle=0,width=\columnwidth]{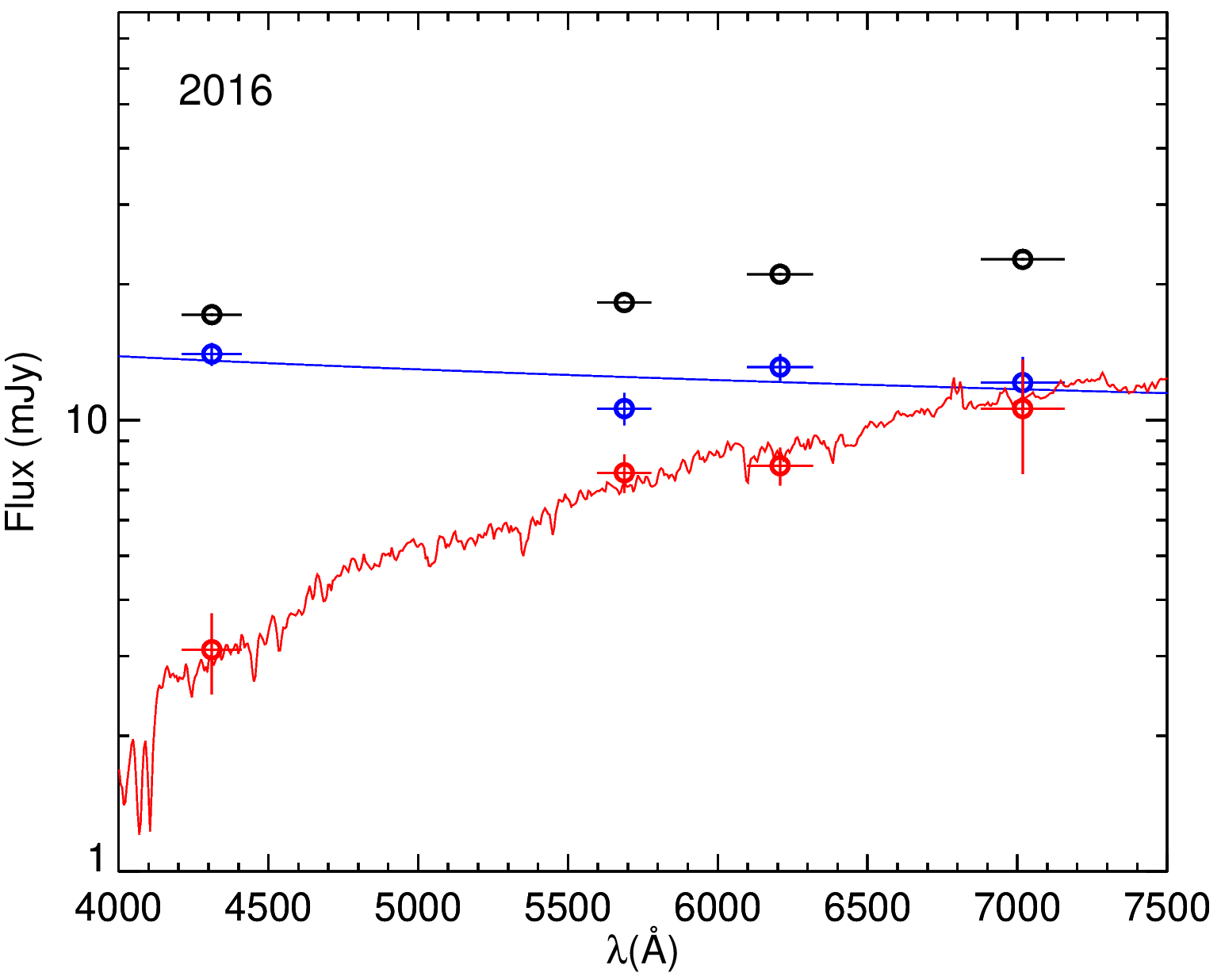}
  \includegraphics[angle=0,width=\columnwidth]{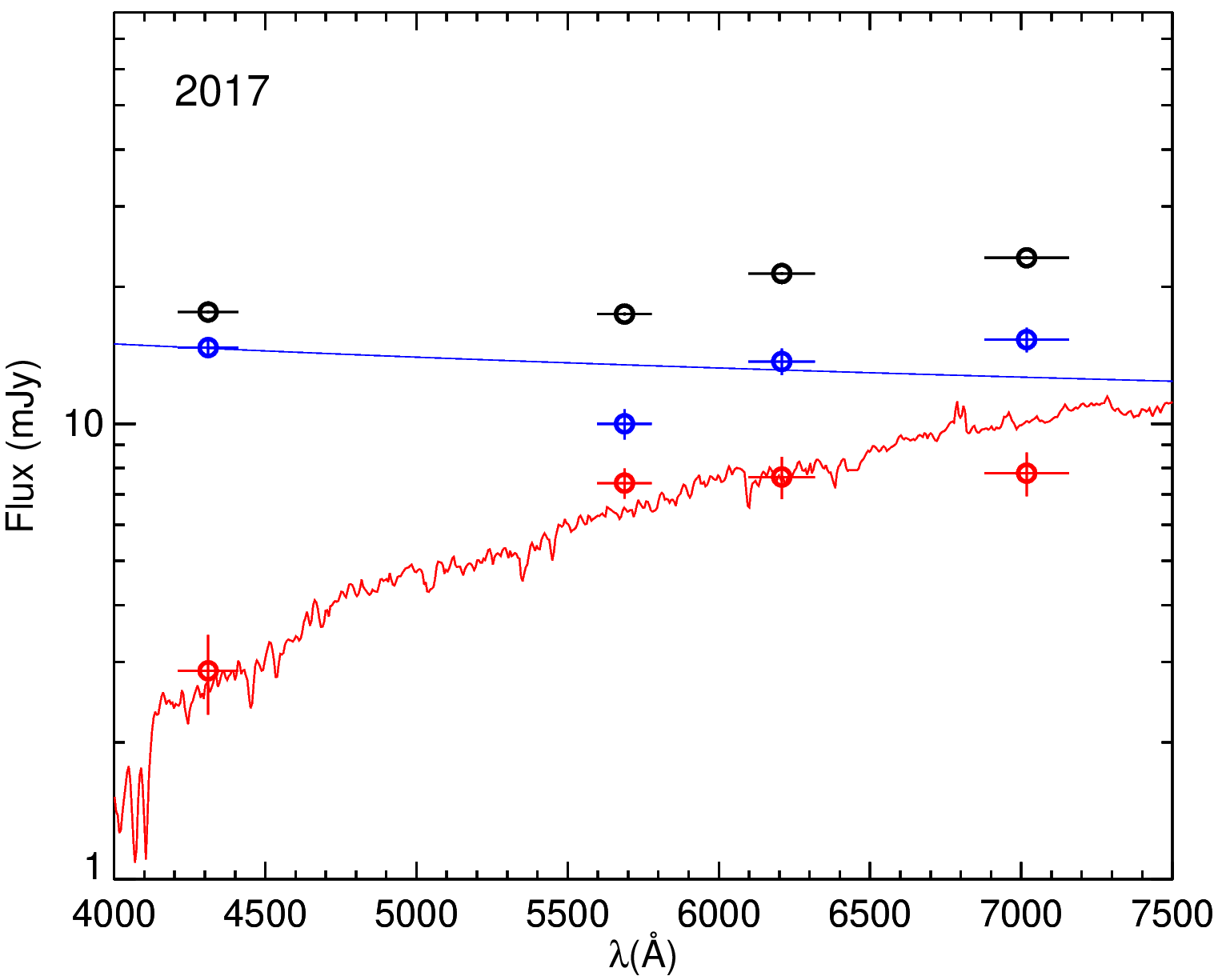}
  \caption{Flux decomposition as obtained from the FVG analysis for epochs 2016 (top) and 2017 (bottom). The total fluxes are shown with black circles. The host galaxy subtracted AGN spectrum (blue circles) is consistent with a blue AGN spectrum and follows the accretion disk \citealt{1973A&A....24..337S} model (F$_{\nu}\propto\nu^{1/3}$; blue line). A bulge model template of \citealt{1996ApJ...467...38K} (red line) is a good fit to the host galaxy fluxes derived by the FVG analysis (red circles).}
  \label{sed}
\end{figure*}

\begin{figure*}
\includegraphics[width=0.75\columnwidth]{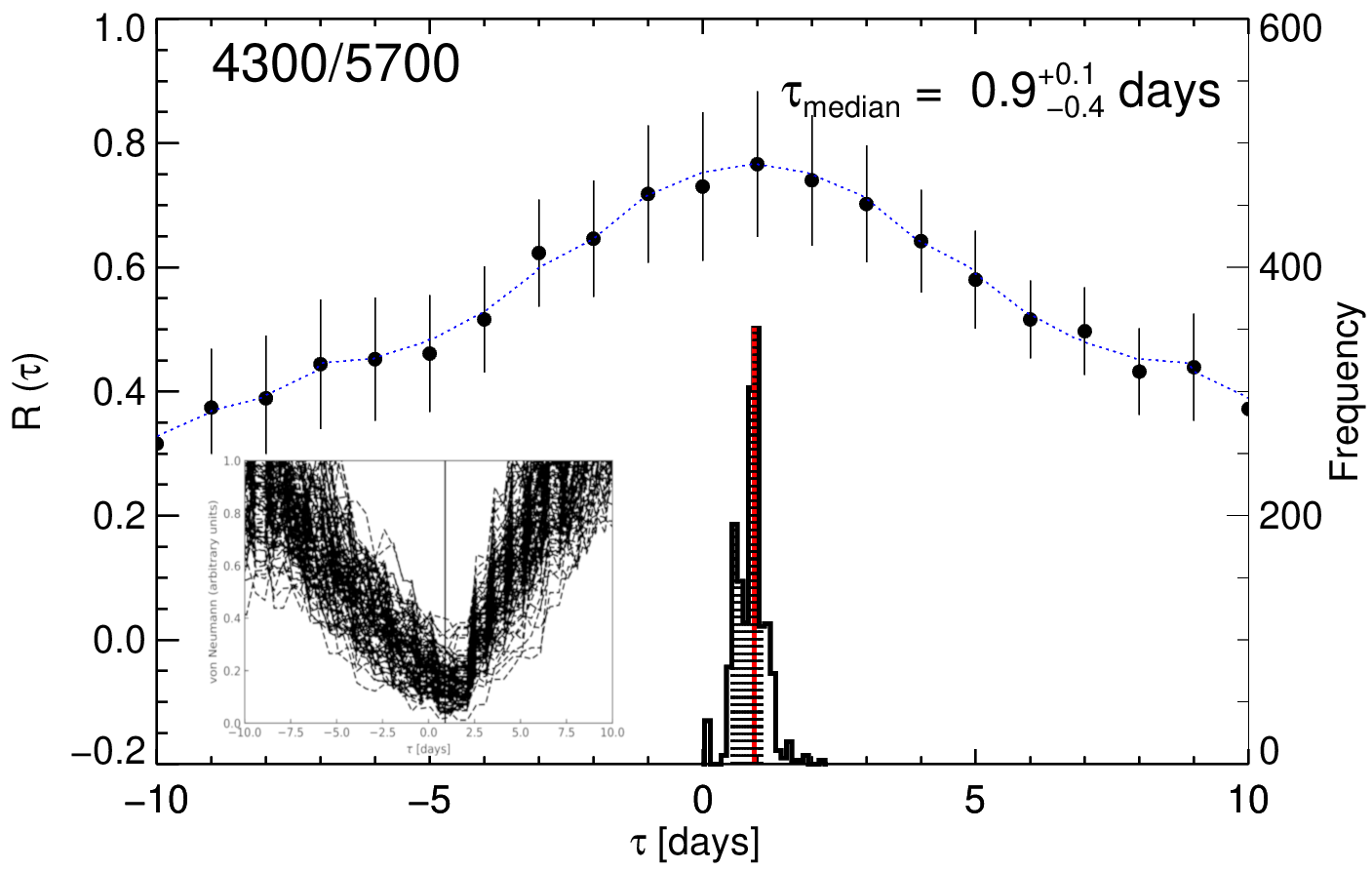} 
\includegraphics[width=0.75\columnwidth]{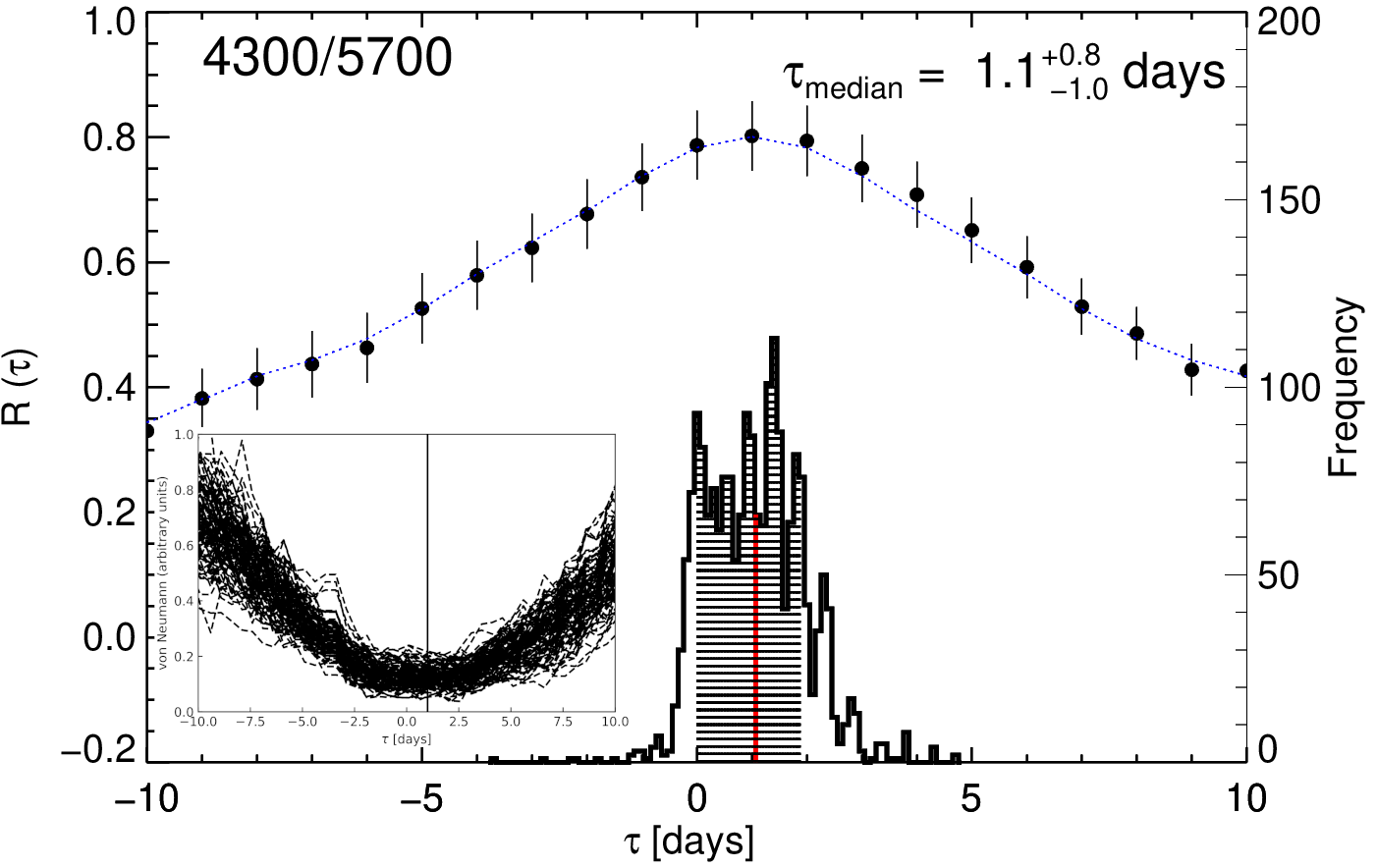} 
\includegraphics[width=0.75\columnwidth]{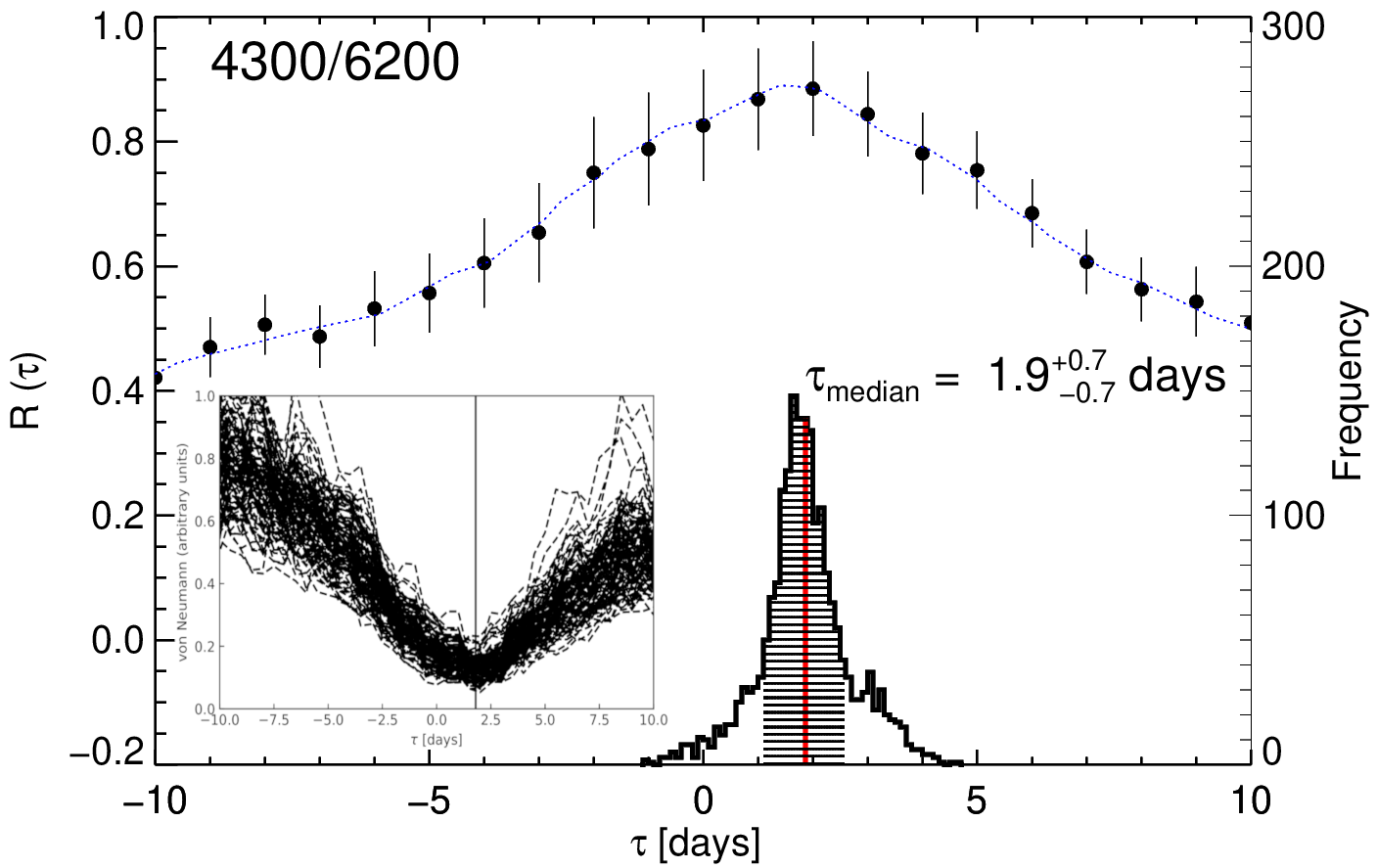}
\includegraphics[width=0.75\columnwidth]{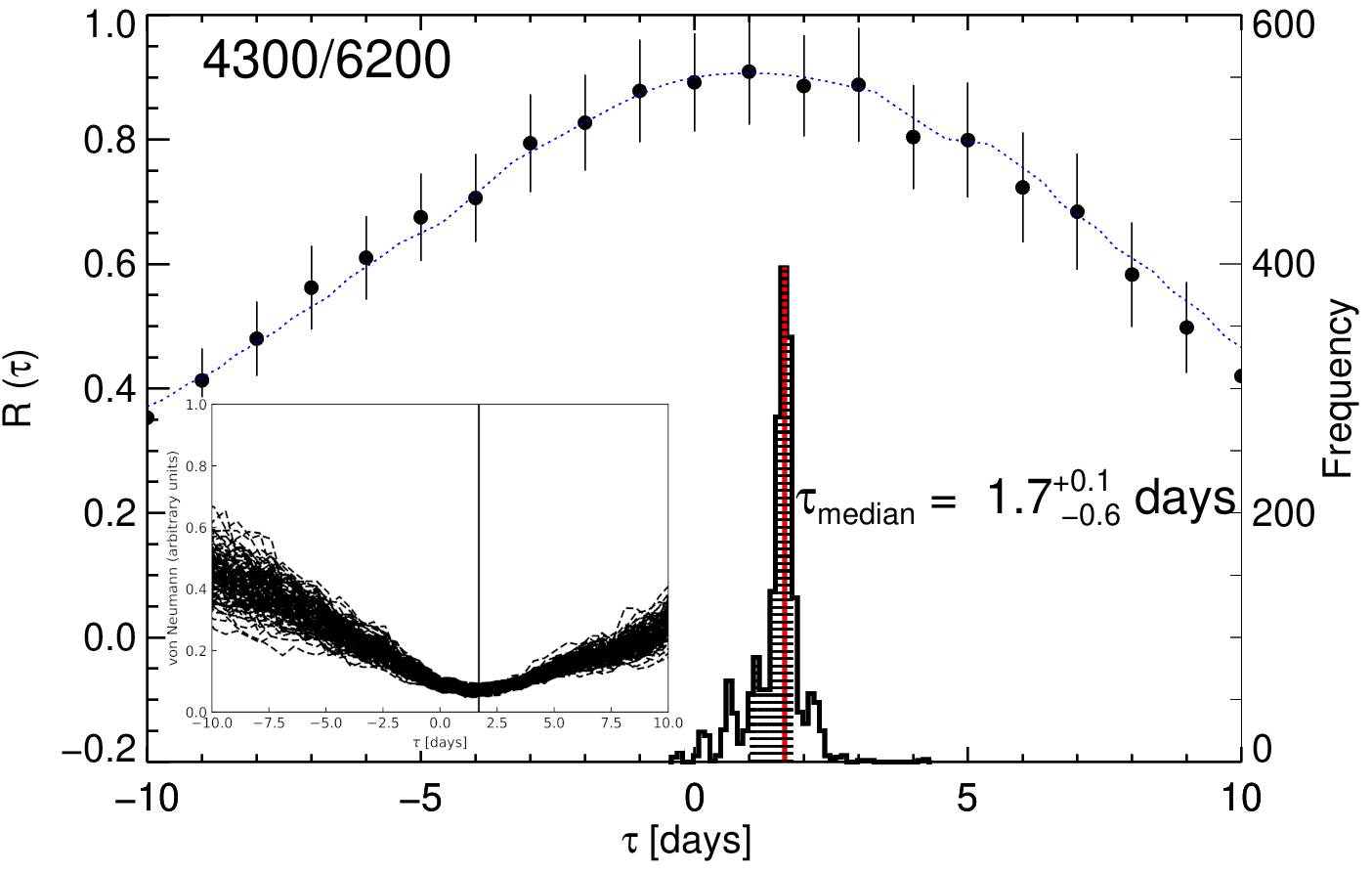} 
\includegraphics[width=0.75\columnwidth]{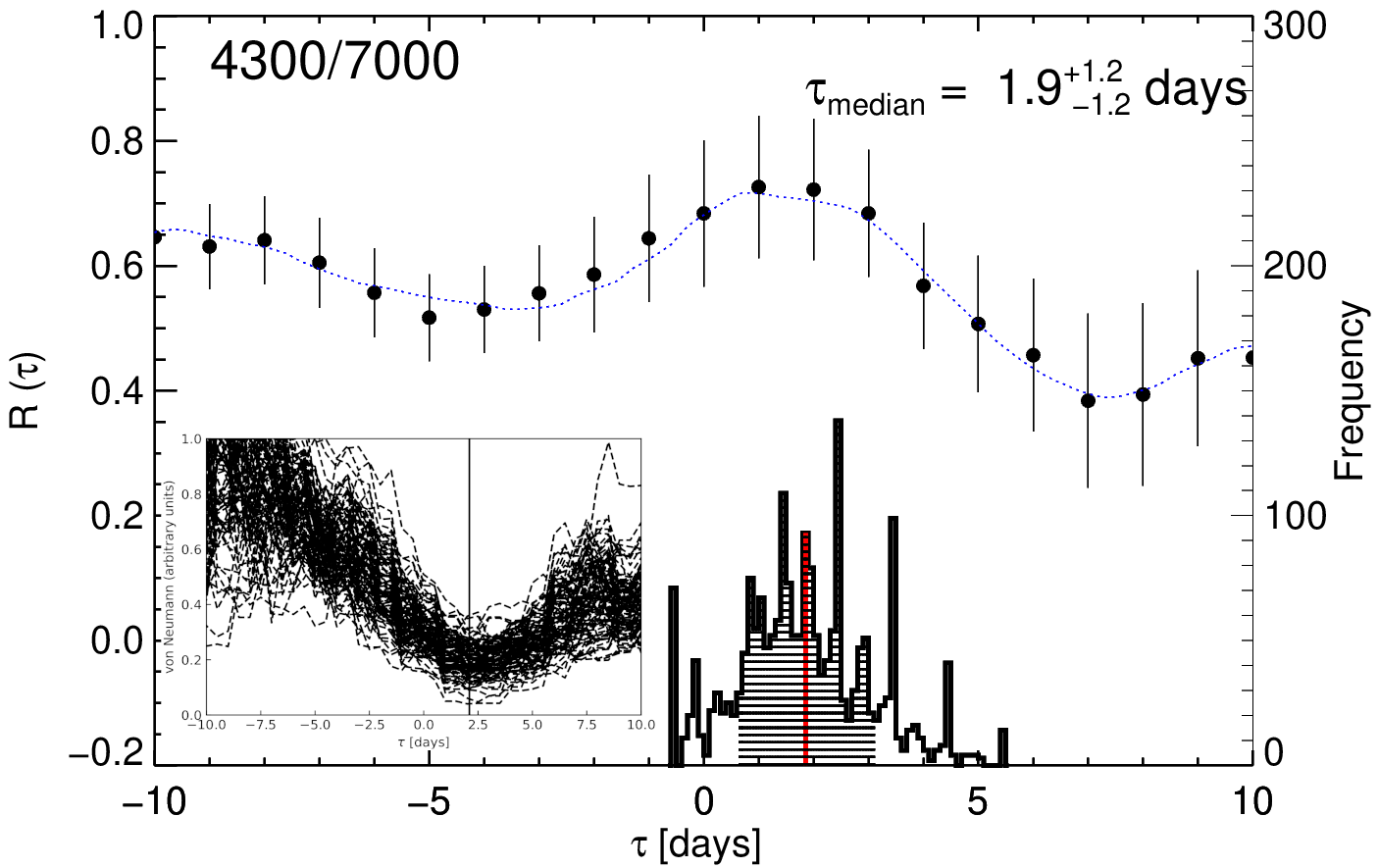} 
\includegraphics[width=0.75\columnwidth]{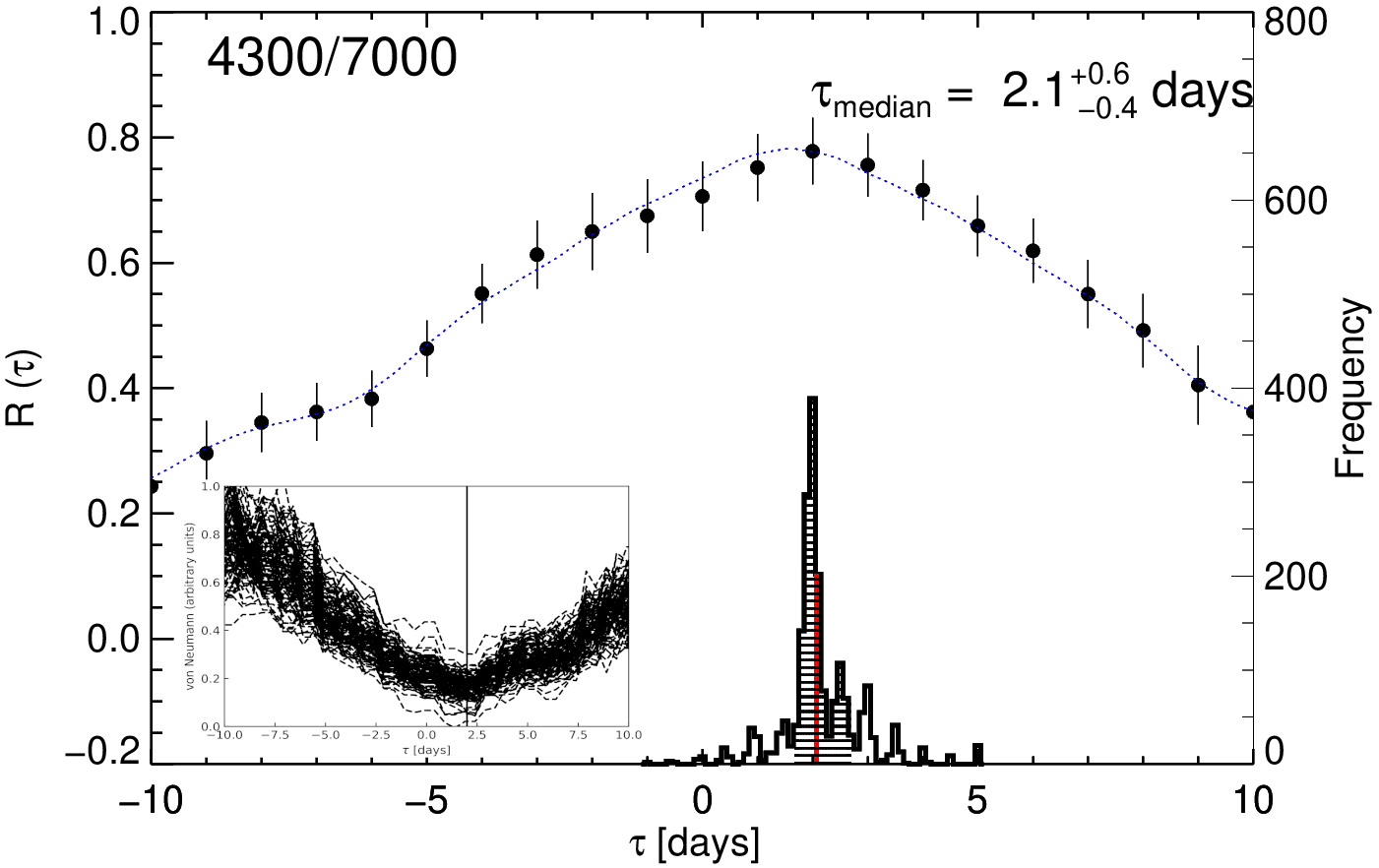}
\caption{Results of the time delay analysis for 2016 (left) and 2017 (right) campaigns. The interpolated cross correlation function (ICCF) is shown as blue dotted lines, while the discrete correlation function (DCF) is shown as black circles with $\pm 1\sigma$ error bars. The histograms shows the distribution of the centroid time delay obtained by cross correlating 2000 flux randomized and randomly selected subset light curves (FR/RSS method). The black area marks the 68\% confidence range used to calculate the errors of the centroid (red line). The inset shows the von Neumann (VN) estimator obtained for 2000 FR/RSS subset light curves.}
\label{delayccf}
\end{figure*}

\begin{figure*}
  \centering
  \includegraphics[width=\columnwidth]{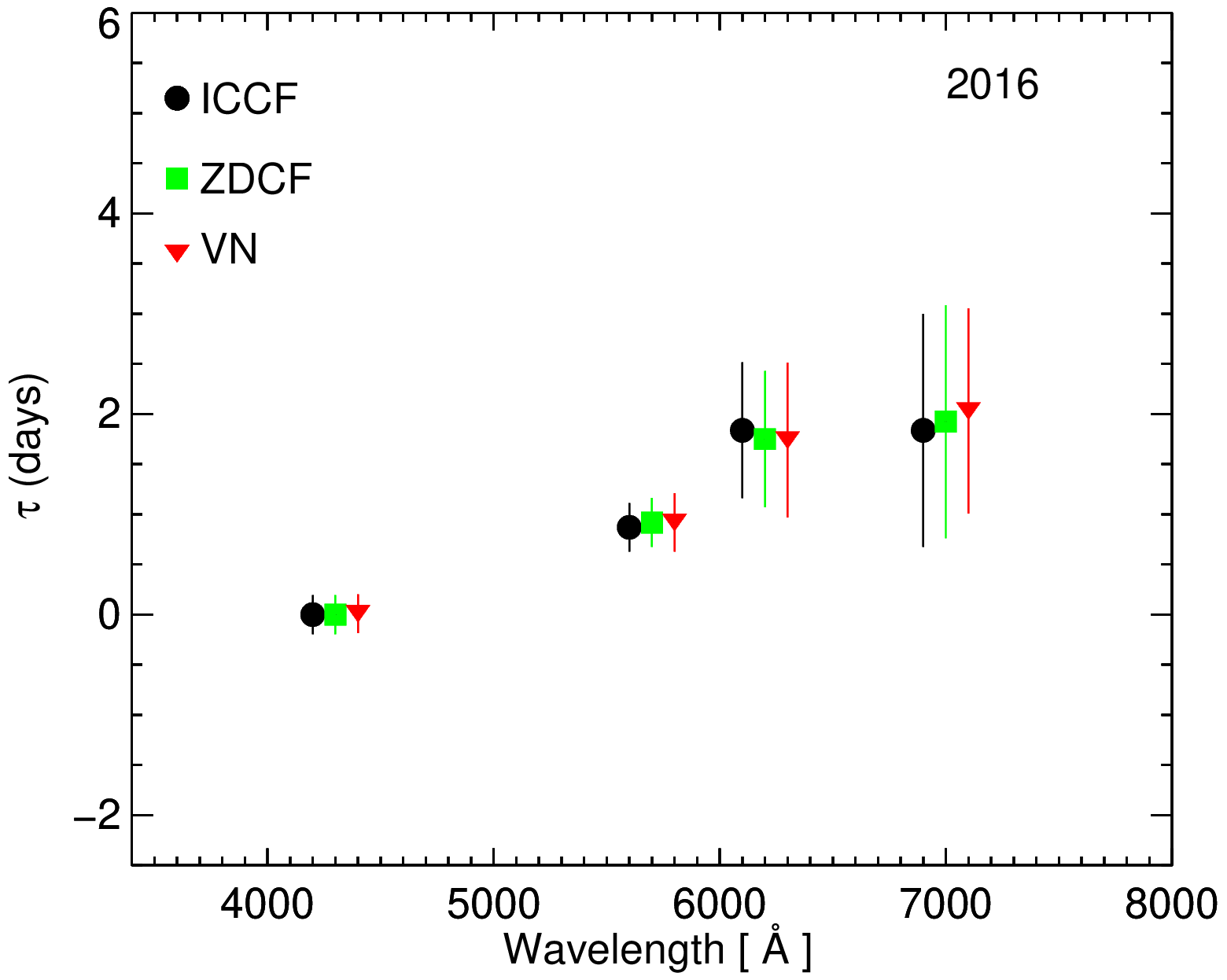}
  \includegraphics[width=\columnwidth]{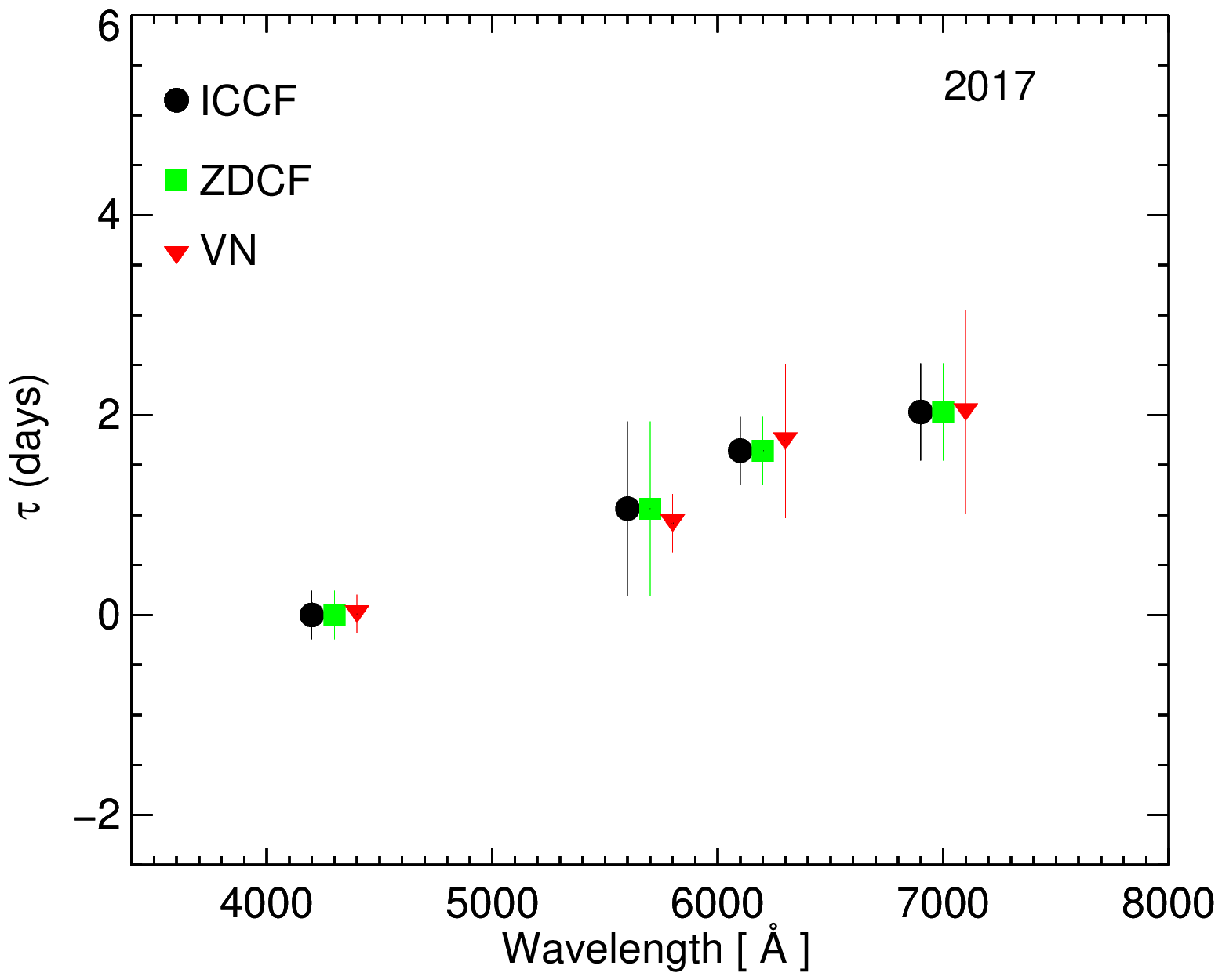}
  \caption{Time delay as a function of wavelength obtained for campaigns 2016 ({\it left}) and 2017 ({\it right}) using the ICCF (black circles), DCF (green squares) and von Neumann (red triangles) methods. The time delays are calculated with respect to the 4300\,\AA\ narrow-band and are corrected by the time dilation factor ($1+z = 1.0344$).}
\label{delayres}
\end{figure*}

\onecolumn

\begin{table*}
\begin{center}
\caption{NB4300, NB5700, NB6200 and NB7000 fluxes corrected by extinction for 2016 campaign. MJD correspond to the modified Julian Date (JD) JD-2,450,000. The fluxes are given in mJy.}
\label{table4}
\begin{tabular}{cccccccc}
\hline
MJD & $F_{4300}$ & MJD & $F_{5700}$ & MJD & $F_{6200}$ & MJD & $F_{7000}$\\ 
\hline
\hline
57547.957 &  $17.222\pm 0.154$ & 57630.738 & $18.135\pm  0.203$ & 57525.035  & $20.712\pm 0.168$ & 57645.715 & $22.376\pm  0.250$ \\
57551.008 &  $17.301\pm 0.137$ & 57632.754 & $17.932\pm  0.092$ & 57542.984  & $20.270\pm 0.210$ & 57646.715 & $22.285\pm  0.204$ \\
57551.961 &  $17.239\pm 0.120$ & 57633.738 & $17.895\pm  0.092$ & 57547.961  & $20.270\pm 0.210$ & 57647.715 & $22.331\pm  0.159$ \\
57552.961 &  $17.154\pm 0.103$ & 57634.777 & $17.950\pm  0.092$ & 57551.016  & $21.112\pm 0.105$ & 57648.711 & $22.376\pm  0.113$ \\
57553.945 &  $17.085\pm 0.103$ & 57635.781 & $17.913\pm  0.092$ & 57551.965  & $21.069\pm 0.105$ & 57650.738 & $22.716\pm  0.113$ \\
57555.941 &  $16.965\pm 0.103$ & 57636.770 & $17.766\pm  0.092$ & 57552.969  & $20.901\pm 0.105$ & 57652.746 & $23.057\pm  0.113$ \\
57556.938 &  $16.897\pm 0.103$ & 57637.746 & $17.581\pm  0.092$ & 57553.953  & $20.901\pm 0.105$ & 57653.738 & $23.080\pm  0.113$ \\
57557.934 &  $16.811\pm 0.103$ & 57639.727 & $17.470\pm  0.092$ & 57555.949  & $20.754\pm 0.105$ & 57654.707 & $23.238\pm  0.113$ \\
57558.938 &  $16.863\pm 0.120$ & 57643.711 & $17.230\pm  0.166$ & 57556.941  & $20.859\pm 0.105$ & 57655.785 & $22.943\pm  0.113$ \\
57559.941 &  $16.777\pm 0.154$ & 57645.711 & $16.990\pm  0.203$ & 57557.938  & $20.775\pm 0.105$ & 57658.703 & $22.989\pm  0.113$ \\
57561.914 &  $16.777\pm 0.137$ & 57646.711 & $17.378\pm  0.185$ & 57558.945  & $20.754\pm 0.105$ & 57659.703 & $22.989\pm  0.113$ \\
57562.973 &  $16.897\pm 0.137$ & 57647.711 & $17.766\pm  0.148$ & 57559.945  & $20.691\pm 0.168$ & 57660.703 & $23.125\pm  0.113$ \\
57563.953 &  $16.811\pm 0.103$ & 57648.707 & $18.061\pm  0.111$ & 57561.922  & $20.565\pm 0.252$ & 57661.703 & $22.921\pm  0.113$ \\
57564.945 &  $16.948\pm 0.103$ & 57650.730 & $18.190\pm  0.092$ & 57562.980  & $20.607\pm 0.252$ & 57662.699 & $22.807\pm  0.113$ \\
57565.922 &  $16.948\pm 0.103$ & 57652.738 & $18.356\pm  0.092$ & 57563.961  & $20.607\pm 0.189$ & 57663.699 & $22.558\pm  0.113$ \\
57567.961 &  $17.410\pm 0.103$ & 57653.734 & $18.320\pm  0.092$ & 57564.953  & $20.922\pm 0.105$ & 57664.746 & $22.626\pm  0.113$ \\
57568.895 &  $17.359\pm 0.086$ & 57654.703 & $18.227\pm  0.092$ & 57565.930  & $20.943\pm 0.105$ & 57665.699 & $22.648\pm  0.113$ \\
57569.949 &  $17.427\pm 0.086$ & 57655.777 & $18.116\pm  0.092$ & 57567.969  & $21.343\pm 0.105$ & 57666.699 & $22.580\pm  0.113$ \\
57570.922 &  $17.222\pm 0.086$ & 57658.699 & $18.227\pm  0.092$ & 57568.902  & $21.448\pm 0.105$ & 57667.738 & $22.421\pm  0.136$ \\
57571.930 &  $17.188\pm 0.086$ & 57659.699 & $18.356\pm  0.092$ & 57569.953  & $21.700\pm 0.105$ & 57668.723 & $22.308\pm  0.113$ \\
57572.902 &  $17.034\pm 0.086$ & 57660.695 & $18.024\pm  0.092$ & 57570.930  & $21.763\pm 0.105$ & 57670.750 & $22.376\pm  0.159$ \\
57574.891 &  $16.897\pm 0.086$ & 57661.695 & $18.135\pm  0.092$ & 57571.938  & $21.742\pm 0.105$ & 57672.695 & $22.376\pm  0.182$ \\
57575.922 &  $16.914\pm 0.086$ & 57662.695 & $18.301\pm  0.092$ & 57572.910  & $21.532\pm 0.105$ & 57673.723 & $22.648\pm  0.159$ \\
57576.895 &  $16.743\pm 0.086$ & 57663.695 & $18.209\pm  0.111$ & 57574.895  & $20.880\pm 0.105$ & 57674.691 & $22.489\pm  0.136$ \\
57578.891 &  $17.017\pm 0.103$ & 57664.742 & $18.098\pm  0.092$ & 57575.930  & $20.775\pm 0.105$ & 57675.691 & $22.830\pm  0.136$ \\
57579.902 &  $17.136\pm 0.103$ & 57665.691 & $18.320\pm  0.092$ & 57576.902  & $20.880\pm 0.105$ & 57677.703 & $22.853\pm  0.113$ \\
57580.895 &  $17.496\pm 0.103$ & 57666.695 & $18.393\pm  0.092$ & 57578.898  & $21.090\pm 0.105$ & 57679.688 & $22.603\pm  0.113$ \\
57581.906 &  $17.598\pm 0.103$ & 57667.734 & $18.597\pm  0.092$ & 57579.906  & $21.217\pm 0.105$ & 57680.691 & $22.580\pm  0.136$ \\
57582.871 &  $17.598\pm 0.103$ & 57668.715 & $18.338\pm  0.111$ & 57580.902  & $21.364\pm 0.105$ & 57681.688 & $22.603\pm  0.113$ \\
57583.941 &  $17.667\pm 0.103$ & 57670.746 & $18.338\pm  0.148$ & 57581.914  & $21.469\pm 0.105$ & 57683.684 & $22.648\pm  0.113$ \\
57584.871 &  $17.547\pm 0.086$ & 57672.688 & $18.560\pm  0.166$ & 57582.879  & $21.532\pm 0.105$ & 57684.684 & $22.853\pm  0.113$ \\
57585.895 &  $17.615\pm 0.086$ & 57673.719 & $18.652\pm  0.148$ & 57583.949  & $21.532\pm 0.126$ & 57687.684 & $22.966\pm  0.159$ \\
57588.883 &  $17.701\pm 0.103$ & 57674.688 & $18.615\pm  0.148$ & 57584.879  & $21.532\pm 0.126$ & -         & -                  \\
57591.844 &  $17.906\pm 0.086$ & 57675.688 & $18.615\pm  0.148$ & 57585.902  & $21.532\pm 0.189$ & -         & -                  \\
57593.844 &  $17.838\pm 0.034$ & 57677.699 & $18.597\pm  0.111$ & 57588.891  & $21.574\pm 0.210$ & -         & -                  \\
57594.891 &  $17.957\pm 0.120$ & 57679.684 & $18.726\pm  0.092$ & 57591.852  & $21.763\pm 0.189$ & -         & -                  \\
57595.840 &  $17.974\pm 0.137$ & 57680.688 & $18.763\pm  0.092$ & 57593.848  & $21.932\pm 0.126$ & -         & -                  \\
57596.848 &  $18.265\pm 0.137$ & 57681.680 & $18.855\pm  0.092$ & 57594.895  & $22.100\pm 0.105$ & -         & -                  \\
57597.844 &  $18.573\pm 0.103$ & 57683.680 & $18.763\pm  0.092$ & 57595.848  & $22.247\pm 0.105$ & -         & -                  \\
57599.832 &  $18.556\pm 0.103$ & 57684.680 & $18.597\pm  0.092$ & 57597.848  & $22.394\pm 0.105$ & -         & -                  \\
57600.816 &  $18.351\pm 0.103$ & 57685.680 & $18.800\pm  0.092$ & 57599.840  & $22.331\pm 0.105$ & -         & -                  \\
57602.812 &  $17.752\pm 0.137$ & 57687.680 & $18.818\pm  0.148$ & 57600.824  & $22.058\pm 0.105$ & -         & -                  \\
57603.855 &  $17.359\pm 0.103$ & -         & -                  & 57602.816  & $21.826\pm 0.105$ & -         & -                  \\
57604.809 &  $17.188\pm 0.137$ & -         & -                  & 57603.863  & $21.574\pm 0.105$ & -         & -                  \\
57605.797 &  $16.863\pm 0.137$ & -         & -                  & 57604.816  & $21.406\pm 0.105$ & -         & -                  \\
57611.820 &  $16.640\pm 0.137$ & -         & -                  & 57605.805  & $21.006\pm 0.105$ & -         & -                  \\
57616.750 &  $16.384\pm 0.137$ & -         & -                  & 57611.828  & $20.838\pm 0.126$ & -         & -                  \\
57626.770 &  $16.658\pm 0.103$ & -         & -                  & 57616.758  & $20.397\pm 0.147$ & -         & -                  \\
57630.727 &  $16.640\pm 0.103$ & -         & -                  & 57630.734  & $20.670\pm 0.126$ & -         & -                  \\
57632.742 &  $16.623\pm 0.103$ & -         & -                  & 57632.750  & $20.586\pm 0.105$ & -         & -                  \\
57633.727 &  $16.487\pm 0.103$ & -         & -                  & 57633.734  & $20.670\pm 0.105$ & -         & -                  \\
57634.766 &  $16.452\pm 0.103$ & -         & -                  & 57634.773  & $20.586\pm 0.105$ & -         & -                  \\
57635.770 &  $16.316\pm 0.103$ & -         & -                  & 57635.777  & $20.397\pm 0.105$ & -         & -                  \\
57636.762 &  $16.179\pm 0.103$ & -         & -                  & 57636.766  & $20.334\pm 0.105$ & -         & -                  \\
57637.734 &  $16.110\pm 0.137$ & -         & -                  & 57637.742  & $20.102\pm 0.105$ & -         & -                  \\
57639.715 &  $15.854\pm 0.086$ & -         & -                  & 57639.719  & $19.976\pm 0.105$ & -         & -                  \\
57643.703 &  $15.649\pm 0.086$ & -         & -                  & 57643.707  & $19.598\pm 0.126$ & -         & -                  \\
57645.703 &  $15.238\pm 0.086$ & -         & -                  & 57645.707  & $19.429\pm 0.168$ & -         & -                  \\
57646.699 &  $15.306\pm 0.086$ & -         & -                  & 57646.707  & $19.577\pm 0.168$ & -         & -                  \\
57647.699 &  $15.546\pm 0.086$ & -         & -                  & 57647.703  & $20.144\pm 0.126$ & -         & -                  \\
57648.695 &  $16.008\pm 0.120$ & -         & -                  & 57648.703  & $20.207\pm 0.105$ & -         & -                  \\
\hline                                       
\end{tabular}
\end{center}
\end{table*}


\begin{table*}
\begin{center}
\contcaption{}
\begin{tabular}{cccccccc}
\hline
MJD & $F_{4300}$ & MJD & $F_{5700}$ & MJD & $F_{6200}$ & MJD & $F_{7000}$\\ 
\hline 
\hline
57650.719 &  $16.298\pm 0.120$ & -         & -                  & 57650.727  & $20.186\pm 0.126$ & -         & -                  \\
57652.727 &  $16.418\pm 0.120$ & -         & -                  & 57652.734  & $19.997\pm 0.126$ & -         & -                  \\
57653.723 &  $16.401\pm 0.120$ & -         & -                  & 57653.730  & $19.976\pm 0.126$ & -         & -                  \\
57654.691 &  $16.264\pm 0.103$ & -         & -                  & 57654.699  & $19.850\pm 0.126$ & -         & -                  \\
57655.770 &  $16.247\pm 0.103$ & -         & -                  & 57655.773  & $19.892\pm 0.105$ & -         & -                  \\
57658.688 &  $16.367\pm 0.103$ & -         & -                  & 57658.695  & $20.249\pm 0.105$ & -         & -                  \\
57659.688 &  $16.469\pm 0.120$ & -         & -                  & 57659.695  & $20.565\pm 0.105$ & -         & -                  \\
57660.688 &  $16.675\pm 0.120$ & -         & -                  & 57660.691  & $20.544\pm 0.126$ & -         & -                  \\
57661.688 &  $16.623\pm 0.120$ & -         & -                  & 57661.691  & $20.523\pm 0.147$ & -         & -                  \\
57662.684 &  $16.743\pm 0.120$ & -         & -                  & 57662.691  & $20.481\pm 0.105$ & -         & -                  \\
57663.684 &  $16.640\pm 0.120$ & -         & -                  & 57663.691  & $20.565\pm 0.105$ & -         & -                  \\
57664.684 &  $16.743\pm 0.120$ & -         & -                  & 57664.691  & $20.565\pm 0.126$ & -         & -                  \\
57665.684 &  $16.777\pm 0.103$ & -         & -                  & 57665.688  & $20.733\pm 0.126$ & -         & -                  \\
57666.684 &  $16.709\pm 0.103$ & -         & -                  & 57666.691  & $20.586\pm 0.105$ & -         & -                  \\
57667.723 &  $16.829\pm 0.120$ & -         & -                  & 57667.727  & $20.523\pm 0.126$ & -         & -                  \\
57668.707 &  $16.829\pm 0.120$ & -         & -                  & 57668.711  & $20.565\pm 0.105$ & -         & -                  \\
57670.734 &  $16.914\pm 0.120$ & -         & -                  & 57670.742  & $20.586\pm 0.126$ & -         & -                  \\
57672.680 &  $16.931\pm 0.103$ & -         & -                  & 57672.684  & $20.712\pm 0.147$ & -         & -                  \\
57673.707 &  $17.154\pm 0.103$ & -         & -                  & 57673.711  & $20.670\pm 0.126$ & -         & -                  \\
57674.676 &  $17.256\pm 0.103$ & -         & -                  & 57674.680  & $20.859\pm 0.105$ & -         & -                  \\
57675.676 &  $17.188\pm 0.103$ & -         & -                  & 57675.680  & $20.859\pm 0.105$ & -         & -                  \\
57677.688 &  $17.290\pm 0.120$ & -         & -                  & 57677.691  & $21.112\pm 0.105$ & -         & -                  \\
57678.672 &  $17.171\pm 0.120$ & -         & -                  & 57678.680  & $20.985\pm 0.105$ & -         & -                  \\
57679.672 &  $17.188\pm 0.103$ & -         & -                  & 57679.680  & $21.154\pm 0.126$ & -         & -                  \\
57680.676 &  $17.239\pm 0.120$ & -         & -                  & 57680.684  & $21.006\pm 0.105$ & -         & -                  \\
57681.672 &  $17.564\pm 0.103$ & -         & -                  & 57681.676  & $21.427\pm 0.105$ & -         & -                  \\
57683.668 &  $17.188\pm 0.120$ & -         & -                  & 57683.676  & $21.553\pm 0.126$ & -         & -                  \\
57684.668 &  $17.068\pm 0.103$ & -         & -                  & 57684.676  & $21.427\pm 0.105$ & -         & -                  \\
57685.668 &  $17.273\pm 0.103$ & -         & -                  & 57685.676  & $21.658\pm 0.126$ & -         & -                  \\
57692.703 &  $18.231\pm 0.120$ & -         & -                  & 57687.672  & $21.995\pm 0.105$ & -         & -                  \\
57694.664 &  $18.847\pm 0.120$ & -         & -                  & 57692.711  & $22.478\pm 0.105$ & -         & -                  \\
57695.699 &  $19.018\pm 0.120$ & -         & -                  & 57694.672  & $22.752\pm 0.126$ & -         & -                  \\
57696.660 &  $19.086\pm 0.120$ & -         & -                  & 57695.707  & $22.647\pm 0.105$ & -         & -                  \\
57698.660 &  $19.497\pm 0.120$ & -         & -                  & 57696.668  & $22.562\pm 0.105$ & -         & -                  \\
57699.664 &  $20.146\pm 0.103$ & -         & -                  & 57698.668  & $22.668\pm 0.126$ & -         & -                  \\
-         &  -     	       & -         & -                  & 57699.672  & $23.109\pm 0.126$ & -         & -                  \\
\hline                                       
\end{tabular}
\end{center}
\end{table*}


\begin{table*}
\begin{center}
\caption{The same as Table A1 but for 2017 campaign.}
\label{2017_fluxes}
\begin{tabular}{cccccccc}
\hline 
MJD & $F_{4300}$ & MJD & $F_{5700}$ & MJD & $F_{6200}$ & MJD & $F_{7000}$\\ 
\hline
\hline
57900.988 &  $21.978\pm 0.123$ & 57917.930 & $17.643\pm  0.1923$ & 57900.996  & $24.053\pm 0.128$ & 57965.840 & $22.311\pm  0.114$ \\
57904.961 &  $20.713\pm 0.105$ & 57920.992 & $18.062\pm  0.0872$ & 57904.969  & $24.117\pm 0.107$ & 57966.820 & $22.129\pm  0.114$ \\
57907.008 &  $20.362\pm 0.088$ & 57921.957 & $17.643\pm  0.0872$ & 57907.016  & $24.288\pm 0.107$ & 57967.848 & $22.220\pm  0.114$ \\
57907.988 &  $20.327\pm 0.105$ & 57922.938 & $17.818\pm  0.0872$ & 57907.996  & $23.903\pm 0.107$ & 57968.832 & $22.152\pm  0.114$ \\
57909.004 &  $19.923\pm 0.105$ & 57923.953 & $18.027\pm  0.0872$ & 57909.012  & $23.689\pm 0.107$ & 57969.828 & $22.107\pm  0.136$ \\
57910.004 &  $19.413\pm 0.088$ & 57924.953 & $17.922\pm  0.0872$ & 57910.012  & $23.561\pm 0.107$ & 57970.844 & $22.311\pm  0.227$ \\
57910.934 &  $19.009\pm 0.123$ & 57925.926 & $17.853\pm  0.0872$ & 57910.945  & $22.984\pm 0.128$ & 57971.836 & $22.107\pm  0.114$ \\
57911.934 &  $18.746\pm 0.088$ & 57929.914 & $17.591\pm  0.0872$ & 57911.941  & $22.599\pm 0.107$ & 57973.852 & $22.629\pm  0.159$ \\
57913.012 &  $18.113\pm 0.123$ & 57930.930 & $17.713\pm  0.0876$ & 57913.020  & $22.150\pm 0.171$ & 57974.809 & $22.993\pm  0.136$ \\
57915.957 &  $17.164\pm 0.123$ & 57931.934 & $17.818\pm  0.0873$ & 57917.922  & $21.231\pm 0.214$ & 57975.801 & $23.288\pm  0.136$ \\
57917.914 &  $17.358\pm 0.123$ & 57932.906 & $17.818\pm  0.0875$ & 57920.988  & $21.530\pm 0.107$ & 57976.828 & $23.220\pm  0.136$ \\
57920.980 &  $17.867\pm 0.088$ & 57933.938 & $17.870\pm  0.0878$ & 57921.953  & $21.102\pm 0.107$ & 57977.840 & $23.311\pm  0.136$ \\
57921.941 &  $17.358\pm 0.105$ & 57935.895 & $17.382\pm  0.0701$ & 57922.934  & $21.209\pm 0.107$ & 57978.883 & $23.743\pm  0.136$ \\
57922.922 &  $17.463\pm 0.088$ & 57936.965 & $17.382\pm  0.0872$ & 57923.949  & $21.252\pm 0.107$ & 57979.852 & $23.834\pm  0.136$ \\
57923.941 &  $17.727\pm 0.088$ & 57937.957 & $17.138\pm  0.0872$ & 57924.945  & $21.316\pm 0.107$ & 57982.805 & $23.606\pm  0.136$ \\
57924.938 &  $17.692\pm 0.105$ & 57938.902 & $16.824\pm  0.0872$ & 57925.922  & $21.594\pm 0.107$ & 57983.801 & $23.379\pm  0.114$ \\
57925.910 &  $17.463\pm 0.088$ & 57939.902 & $16.876\pm  0.1572$ & 57929.910  & $20.717\pm 0.107$ & 57987.754 & $23.515\pm  0.136$ \\
57926.895 &  $17.129\pm 0.123$ & 57951.871 & $16.650\pm  0.1052$ & 57930.922  & $20.760\pm 0.107$ & 57988.797 & $23.652\pm  0.114$ \\
57929.902 &  $16.690\pm 0.088$ & 57952.859 & $16.284\pm  0.0872$ & 57931.926  & $20.717\pm 0.107$ & 57989.785 & $23.902\pm  0.114$ \\
57930.914 &  $16.637\pm 0.088$ & 57953.902 & $16.284\pm  0.1222$ & 57932.902  & $20.696\pm 0.107$ & 57990.738 & $24.015\pm  0.159$ \\
57931.918 &  $16.848\pm 0.088$ & 57956.848 & $16.301\pm  0.1222$ & 57933.934  & $20.568\pm 0.107$ & 57991.797 & $24.106\pm  0.114$ \\
57932.895 &  $16.743\pm 0.088$ & 57957.836 & $16.458\pm  0.1052$ & 57935.887  & $20.162\pm 0.107$ & 57992.797 & $24.265\pm  0.136$ \\
57933.926 &  $16.514\pm 0.105$ & 57958.848 & $16.440\pm  0.1222$ & 57936.961  & $19.991\pm 0.107$ & 57993.793 & $24.129\pm  0.114$ \\
57935.879 &  $16.268\pm 0.088$ & 57959.824 & $16.667\pm  0.1391$ & 57937.949  & $19.755\pm 0.107$ & 57995.789 & $23.652\pm  0.136$ \\
57936.953 &  $16.110\pm 0.088$ & 57960.871 & $16.894\pm  0.0872$ & 57938.895  & $19.713\pm 0.107$ & 57996.789 & $23.629\pm  0.136$ \\
57937.941 &  $15.917\pm 0.088$ & 57961.828 & $16.981\pm  0.0872$ & 57939.895  & $19.841\pm 0.171$ & 57997.789 & $23.584\pm  0.227$ \\
57938.887 &  $15.917\pm 0.105$ & 57962.855 & $16.946\pm  0.0872$ & 57951.863  & $19.371\pm 0.107$ & 57999.789 & $23.447\pm  0.136$ \\
57939.887 &  $16.023\pm 0.123$ & 57963.828 & $16.981\pm  0.1052$ & 57952.852  & $19.199\pm 0.128$ & 58000.793 & $23.402\pm  0.114$ \\
57951.855 &  $15.829\pm 0.105$ & 57964.879 & $16.754\pm  0.1221$ & 57953.898  & $19.285\pm 0.128$ & 58003.727 & $23.016\pm  0.159$ \\
57952.844 &  $15.583\pm 0.105$ & 57965.832 & $16.737\pm  0.0878$ & 57956.840  & $19.542\pm 0.128$ & 58004.727 & $22.743\pm  0.114$ \\
57953.891 &  $15.636\pm 0.123$ & 57966.816 & $16.597\pm  0.1056$ & 57957.832  & $19.627\pm 0.128$ & 58006.727 & $22.470\pm  0.136$ \\
57956.832 &  $15.987\pm 0.123$ & 57967.840 & $16.702\pm  0.1228$ & 57958.840  & $19.755\pm 0.107$ & 58007.801 & $22.470\pm  0.136$ \\
57957.824 &  $16.093\pm 0.123$ & 57968.828 & $16.876\pm  0.1058$ & 57959.816  & $19.862\pm 0.107$ & 58008.723 & $22.698\pm  0.136$ \\
57958.832 &  $16.110\pm 0.141$ & 57969.820 & $16.667\pm  0.1058$ & 57960.863  & $19.884\pm 0.107$ & 58009.723 & $22.925\pm  0.114$ \\
57959.809 &  $16.110\pm 0.176$ & 57970.840 & $16.650\pm  0.1741$ & 57961.820  & $20.247\pm 0.128$ & 58016.715 & $22.947\pm  0.114$ \\
57960.855 &  $16.497\pm 0.105$ & 57971.828 & $16.719\pm  0.1392$ & 57962.852  & $20.696\pm 0.128$ & 58017.715 & $22.993\pm  0.114$ \\
57961.812 &  $16.673\pm 0.123$ & 57973.848 & $16.772\pm  0.2092$ & 57963.824  & $20.889\pm 0.128$ & -         & -                  \\
57962.844 &  $16.954\pm 0.123$ & 57974.801 & $16.911\pm  0.1922$ & 57964.875  & $20.675\pm 0.128$ & -         & -                  \\
57963.816 &  $16.813\pm 0.105$ & 57975.797 & $17.120\pm  0.1052$ & 57965.824  & $20.717\pm 0.128$ & -         & -                  \\
57964.863 &  $16.655\pm 0.158$ & 57976.824 & $17.190\pm  0.1572$ & 57966.809  & $20.803\pm 0.150$ & -         & -                  \\
57965.816 &  $16.743\pm 0.105$ & 57977.836 & $17.556\pm  0.1222$ & 57967.836  & $20.782\pm 0.128$ & -         & -                  \\
57966.801 &  $16.813\pm 0.105$ & 57978.875 & $17.748\pm  0.1058$ & 57968.820  & $20.846\pm 0.128$ & -         & -                  \\
57967.828 &  $16.655\pm 0.123$ & 57979.844 & $18.027\pm  0.1220$ & 57969.816  & $20.889\pm 0.150$ & -         & -                  \\
57968.812 &  $16.831\pm 0.123$ & 57982.797 & $17.765\pm  0.1221$ & 57970.832  & $20.995\pm 0.214$ & -         & -                  \\
57969.809 &  $17.042\pm 0.105$ & 57983.793 & $17.905\pm  0.1223$ & 57971.820  & $21.102\pm 0.214$ & -         & -                  \\
57970.824 &  $17.024\pm 0.176$ & 57987.750 & $18.149\pm  0.1222$ & 57973.840  & $21.081\pm 0.192$ & -         & -                  \\
57971.812 &  $17.217\pm 0.176$ & 57988.789 & $18.097\pm  0.1051$ & 57974.797  & $20.974\pm 0.128$ & -         & -                  \\
57973.832 &  $17.551\pm 0.228$ & 57989.781 & $18.323\pm  0.1050$ & 57975.789  & $20.953\pm 0.107$ & -         & -                  \\
57974.789 &  $17.358\pm 0.211$ & 57990.734 & $18.480\pm  0.1053$ & 57976.816  & $21.166\pm 0.107$ & -         & -                  \\
57975.781 &  $17.323\pm 0.105$ & 57991.789 & $18.550\pm  0.1222$ & 57977.828  & $21.487\pm 0.128$ & -         & -                  \\
57976.809 &  $17.815\pm 0.123$ & 57992.789 & $18.777\pm  0.1391$ & 57978.871  & $22.107\pm 0.128$ & -         & -                  \\
57977.820 &  $18.236\pm 0.105$ & 57993.785 & $18.899\pm  0.1222$ & 57979.840  & $22.086\pm 0.128$ & -         & -                  \\
57978.863 &  $18.201\pm 0.105$ & 57995.785 & $18.341\pm  0.1570$ & 57982.789  & $22.064\pm 0.128$ & -         & -                  \\
57979.832 &  $18.377\pm 0.123$ & 57996.785 & $18.271\pm  0.1741$ & 57983.785  & $22.043\pm 0.128$ & -         & -                  \\
57982.781 &  $18.271\pm 0.105$ & 57997.781 & $17.783\pm  0.1741$ & 57987.742  & $21.808\pm 0.107$ & -         & -                  \\
57983.777 &  $18.570\pm 0.123$ & 57999.781 & $17.678\pm  0.1570$ & 57988.785  & $21.979\pm 0.128$ & -         & -                  \\
57987.734 &  $18.658\pm 0.123$ & 58000.785 & $17.556\pm  0.1742$ & 57989.773  & $22.107\pm 0.128$ & -         & -                  \\
57988.777 &  $18.394\pm 0.105$ & 58003.723 & $17.452\pm  0.1573$ & 57990.727  & $22.150\pm 0.107$ & -         & -                  \\
57989.766 &  $18.377\pm 0.123$ & 58004.723 & $17.469\pm  0.2094$ & 57991.781  & $22.364\pm 0.128$ & -         & -                  \\
57990.719 &  $18.658\pm 0.088$ & 58006.719 & $17.417\pm  0.1742$ & 57992.781  & $22.471\pm 0.107$ & -         & -                  \\
57991.773 &  $19.009\pm 0.123$ & 58007.793 & $17.469\pm  0.1051$ & 57993.781  & $22.556\pm 0.107$ & -         & -                  \\
57992.773 &  $19.290\pm 0.088$ & 58008.719 & $17.469\pm  0.1051$ & 57995.777  & $22.193\pm 0.214$ & -         & -                  \\
57993.770 &  $18.904\pm 0.105$ & 58009.715 & $17.609\pm  0.1740$ & 57996.777  & $21.765\pm 0.150$ & -         & -                  \\
57995.770 &  $18.289\pm 0.123$ & 58016.707 & $17.399\pm  0.1052$ & 57997.777  & $21.808\pm 0.214$ & -         & -                  \\
\hline
\end{tabular}
\end{center}
\end{table*}

\begin{table*}
\begin{center}
\contcaption{}
\begin{tabular}{cccccccc}
\hline 
MJD & $F_{4300}$ & MJD & $F_{5700}$ & MJD & $F_{6200}$ & MJD & $F_{7000}$\\ 
\hline 
\hline
57996.770 &  $18.008\pm 0.123$ & 58017.707 & $17.609\pm  0.1051$ & 57999.773  & $21.915\pm 0.128$ & -         & -                  \\
57997.770 &  $18.025\pm 0.176$ & -         & -                   & 58000.777  & $21.765\pm 0.107$ & -         & -                  \\
57999.766 &  $17.920\pm 0.105$ & -         & -                   & 58003.715  & $21.530\pm 0.150$ & -         & -                  \\
58000.770 &  $17.674\pm 0.088$ & -         & -                   & 58004.715  & $21.658\pm 0.128$ & -         & -                  \\
58003.707 &  $17.569\pm 0.105$ & -         & -                   & 58006.711  & $21.765\pm 0.128$ & -         & -                  \\
58004.707 &  $17.692\pm 0.123$ & -         & -                   & 58007.789  & $21.787\pm 0.150$ & -         & -                  \\
58006.703 &  $17.533\pm 0.123$ & -         & -                   & 58008.711  & $21.851\pm 0.150$ & -         & -                  \\
58007.781 &  $17.727\pm 0.141$ & -         & -                   & 58009.711  & $21.872\pm 0.171$ & -         & -                  \\
58008.703 &  $17.744\pm 0.123$ & -         & -                   & 58016.703  & $21.551\pm 0.128$ & -         & -                  \\
58009.703 &  $17.674\pm 0.176$ & -         & -                   & 58017.703  & $21.637\pm 0.128$ & -         & -                  \\
58016.695 &  $17.621\pm 0.123$ & -         & -                   & -          & -                 & -         & -                  \\
58017.695 &  $17.621\pm 0.105$ & -         & -                   & -          & -                 & -         & -                  \\
\hline
\end{tabular}
\end{center}
\end{table*}


\bsp	
\label{lastpage}
\end{document}